\documentclass[aps,pra,reprint,10pt,a4paper,nofootinbib,showpacs,varwidth]{revtex4-1}
\usepackage[colorlinks=true, citecolor=red, urlcolor=blue ]{hyperref}
\usepackage[utf8]{inputenc}
\usepackage{array}
\usepackage{graphics}
\usepackage{epsfig,amssymb,amsfonts,amsmath,amsthm,braket,color,xcolor}
\usepackage[normalem]{ulem}

\newcommand{\stkout}[1]{\ifmmode\text{\sout{\ensuremath{#1}}}\else\sout{#1}\fi}
\begin{document}
\title{Correlation between resource-generating  powers of quantum gates}
\author{Aparajita Bhattacharyya, Ahana Ghoshal, Ujjwal Sen}
\affiliation{
Harish-Chandra Research Institute, A CI of Homi Bhabha National Institute, Chhatnag Road, Jhunsi, Prayagraj 211 019, India}

\begin{abstract}
We analyze the optimal  basis for generating the maximum relative entropy of quantum coherence by an arbitrary gate on a two-qubit system. The optimal basis is not unique, and the high quantum coherence generating gates are also typically high entanglement generating ones and vice versa. However, the profile of the relative frequencies of Haar random unitaries generating different amounts of  entanglement for a fixed amount of quantum coherence is different from the one in which the roles of entanglement and quantum coherence are reversed, although both follow the beta distribution.
\end{abstract}

\maketitle
\section{Introduction}
The characterisation of resource theory, within quantum information science and technology, was initiated with the theory of entanglement~\cite{Plenio1, Horodecki,Toth,Lewenstein}.
In general, a resource theory is constructed by putting certain natural ``constraints'' or restrictions on the set of all quantum mechanical operations to perform a specific job, and the restrictions are overcome by utilizing certain ``resources". The naturality of the constraints of course depend on the accessible physical system and the job at hand. In entanglement theory and practice, the restriction is set by the constraint that the observers - generally assumed to be at distant locations - will be able to perform only local quantum operations and classical communication  (LOCC) and the  ``resources'' are the entangled states shared by the same observers. In this manner, entanglement becomes useful in several interesting phenomena and tasks like quantum teleportation~\cite{Bennett1,Pirandola,Liu}, quantum cryptography~\cite{Gisin,Pirandola1,Portmann}
and quantum dense coding~\cite{Wiesner,Li}.
Quantum coherence~\cite{Plenio,Aberg,Winter,Streltsov} (see also~\cite{Sreetama,Chirag,Ignita}) 
is also one of the principal resources in quantum phenomena and information tasks, 
and along with being the reason for the classic interference phenomena, it is also useful in jobs like 
in quantum-enhanced metrology (see e.g. \cite{Pires,Castellini,Zhang}), quantum algorithms (see e.g. \cite{Hillery,Anand,Shi,Liu1,Sen}),  quantum state discrimination (see e.g. \cite{Xiong,Kim}), etc.
The set of allowed operations are now the so-called incoherent operations, and the resources are the quantum coherent states.

Quantifying the entanglement generated by quantum evolutions has been extensively studied in the 
literature (see e.g. \cite{Zanardi2,Zyczkowski,Dodd,Carvalho,Dur,Almeida,Carvalho1,Konrad,Frowis,Moor,Leifer,Bao,Sanders,Cohen,Bennett_arxiv,Kraus, Chefles, Ishizaka, Ye, Plastino, Zanardi1}). In particular, the two-qubit unitaries that generate 
maximal entanglement 
from zero resource pure input states (i.e., two-qubit pure product states) were identified, among other things, in Ref.~\cite{Kraus}. 
The effect of auxiliaries on the entanglement generating ability was considered in e.g.~\cite{Kraus, Bao,Sanders}. The case of mixed state inputs was considered in e.g.~\cite{Moor,Leifer}. 
Generation of maximal entanglement using global unitaries of arbitrary bipartite dimensions was considered in e.g.~\cite{Cohen}. 
The  asymptotic limit of the entanglement generating
capacity of a bidirectional channel acting on two \(d\)-dimensional systems has also been investigated in e.g.~\cite{Bennett_arxiv}.
The relation of entanglement generation  of two-qubit unitary operators with their distinguishability was uncovered in~\cite{Chefles}.
In Ref.~\cite{Ishizaka},  they found 
two-qubit mixed states whose  
entanglement content cannot be increased by unitary transformations. 

Similar to the 
path followed in entanglement theory,
various aspects of the quantitative theory of quantum coherence have been uncovered~\cite{Plenio,Aberg,Winter,Streltsov}. In particular, different properties of the incoherent and coherence-generating  operations have been identified, and 
maximally coherent states 
have been analyzed.
It is understood that there exists 
interconnections between the resource theories of entanglement and quantum coherence. 
But we remember that while entanglement is basis-independent as long as we stick to local bases, quantum coherence is almost strictly basis-dependent. Interrelations between the two resource theories have been studied in e.g.~\cite{Yao,Plenio4,Bera,Ritsch,Fan,Bera1,Qi,Chin,Vedral,Zhu, batayan, Plenio_Egloff, Chitambar2, Oppenheim_Horodecki, Kraemer}.
In particular, given a global unitary operation on a bipartite system, an arbitrary unentangled input state does not generate entanglement. Also, defining quantum coherence with respect to ``local'' bases leads to a conceptualization of entanglement~\cite{Vedral, batayan}. 
There have also been studies 
on  interrelations between quantum coherence and other resources like ``non-locality''~\cite{Pati1, Wu, Pan}, non-markovianity~\cite{Titas,Pati2,Huang,Cakmak,Pati3,Borji} and quantum discord~\cite{Yadin,Matera,Goswami}. Also, a relationship between classical communication- and entanglement-generation 
of
two-qubit unitary operators was 
found in~\cite{Berry}. In~\cite{Pati}, they begin by considering a thermal state and find the maximum amount of quantum coherence that can be generated when acted upon by a unitary operator. 
In~\cite{Yao}, they have constructed a basis-independent measure of quantum coherence and shown that it is equivalent to quantum discord. 
A quantum coherence-generating power for quantum channels, when it acts on incoherent states, was defined in~\cite{Lin}. 
Quantum coherence-generating power of quantum dephasing processes was considered in~\cite{Zanardi}. 
Parallel to the concepts of distillable entanglement~\cite{Smolin,Rains}
and entanglement cost~\cite{Smolin,Hayden},
``quantum coherence distillation'' and ``quantum coherence cost'' were considered in~\cite{Winter}. 

In this paper, we begin by exploring 
the maximal quantum coherence generating  power of an arbitrary two-qubit unitary gate. The parallel case for maximal  \emph{entanglement} generation was considered in~\cite{Kraus}. We provide a formal definition of the  power, which involves a maximum over all product  bases of the underlying tensor-product Hilbert space. We define the same for arbitrary bases also. 

We then compare the quantum coherence generation with entanglement generation for generic two-qubit unitaries. We find in particular that there exists a general tendency of a randomly chosen unitary to produce high quantum coherence when the entanglement generation is high, and conversely to produce high entanglement when the quantum coherence generation is so. 
We make this statement more precise by considering relative frequencies of randomly chosen unitaries to fall in a chosen region on the entanglement-quantum coherence plane (with finite precision). A gist of these facts appears in Figs.~\ref{fig6},~\ref{ent_bas} and~\ref{new_measure} below.
We believe that this correlation between the different resource generating powers for two-qubit unitaries is potentially useful for further analysis of resources and their generations within quantum technologies, as also to lead to fundamental inter-relations between these resources.

Furthermore, we analyze the relative frequency of a randomly generated unitary to have a certain entanglement and quantum coherence generating  power (with finite precision). We find that the profiles of the relative frequencies for a fixed quantum coherence generating  power and a fixed entanglement generating one are qualitatively similar and follow the beta distribution, but are quantitatively different.


The remainder of the paper is arranged as follows. The relevant information from previous literature is discussed in Sec.~\ref{Sec:2}. This includes the definition of entanglement generating  power and its evaluation for two-qubit unitary gates. We also present here a formal definition of the quantum coherence generating  power of unitary gates. 
In Sec.~\ref{Sec:3}, we present our results on quantum coherence  generating power of two-qubit unitaries. 
We compare the entanglement and quantum coherence generations of several paradigmatic two-qubit gates in Sec.~\ref{Sec:4}. In Sec.~\ref{Sec:5}, we consider the same comparison for Haar uniformly generated two-qubit unitaries. We present the concluding remarks in Sec.~\ref{Sec:6}.

\section{Preliminaries}
\label{Sec:2}
We wish to deal with the resource-generating  power of two-party unitaries, when the resource is either entanglement or quantum coherence. 
Let the two parties be  Alice (\(A\)) and Bob (\(B\)), with the system they possess being defined on the Hilbert space, \(\mathcal{H}_A \otimes \mathcal{H}_B\).

The amount of entanglement generated by applying an arbitrary unitary operator $U_{AB}$  is given by
\begin{equation}
\label{eq:1}
E_g(U_{AB})= \max_{\varrho_{AB}}\left[ E\left(U_{AB}    \varrho_{AB} U_{AB}^\dagger \right) - E\left( \varrho_{AB} \right)\right],
\end{equation}
where the maximization is over all states, \(\varrho_{AB}\),  on \(\mathcal{H}_A \otimes \mathcal{H}_B\),
and where $E$ is a measure of entanglement for two-party quantum systems. One can also choose to consider the more general situation where the evolution acts on locally extended Hilbert spaces, and an additional optimization is performed over all such extensions.  
For pure bipartite states, 
the local von Neumann entropy is a good measure of the 
state's entanglement~\cite{Bennett5}, so that use
\begin{equation}
\label{ent}
E(\ket{\psi})
=-\text{Tr}(\rho_A \log_2 \rho_A)=-\text{Tr}(\rho_B \log_2 \rho_B),
\end{equation}
where $\rho_A$ is the partial trace of $|\psi \rangle \langle \psi |$ over the subsystem \(B\) and similarly for $\rho_B$. In this paper, we  restrict to pure input states and to input states having zero resource. Therefore, the input states are pure product states, so that   
\begin{equation}
\label{sadher-lau}
E_g(U) = \max_{|\psi\rangle \otimes |\phi\rangle 
} S\left(\mbox{Tr}_{A/B}P\left[U_{AB}|\psi\rangle_A \otimes |\phi\rangle_B\right]\right),
\end{equation}
where \(P[|\chi\rangle] = |\chi\rangle \langle \chi|\) and \(S(\sigma) = -\mbox{Tr}(\sigma \log_2 \sigma)\).
Moreover, we will restrict ourselves to two-qubit systems, for which the entanglement-generating  power was considered in Ref.~\cite{Kraus}. 
 
 A conceptually different entanglement quantifier can be considered by using the ``Nielsen-Vidal" entanglement monotones, which characterize transformations between bipartite pure states~\cite{Nielsen,Vidal,Hardy,Jonathan_Plenio,Vidal1}.
 If $\ket{\psi}$ is a pure state representing a bipartite quantum system corresponding to  $\mathbb{C}^n \otimes \mathbb{C}^n$, then its Schmidt decomposition is given by
\begin{equation}
    \ket{\psi}=\sum_{i=1}^n \sqrt{\alpha_i} \ket{i_A i_B},
\end{equation}
with $\{\sqrt{\alpha_i}\}$ being the Schmidt coefficients having $\sum_{i=1}^n \alpha_i=1$, and we further assume that they are ordered as $\alpha_i \ge \alpha_{i+1} \ge 0$. $\{\ket{i_A }\}$ and  $\{\ket{i_B }\}$ are the sets of eigenvectors of the reduced subsystems of $\ket{\psi}$, corresponding to the eigenvalues $\alpha_i$. A family of entanglement monotones, $E_k(\ket{\psi})$, for a positive bipartite system, for $k=1,2,...,n$, can be defined 
as~\cite{Vidal}
\begin{equation}
    E_k(\ket{\psi})=\sum_{i=k}^n \alpha_i.
\end{equation}
So, the amount of entanglement generated by applying an arbitrary unitary operator $U_{AB}$, by using $E_k(\ket{\psi})$ as a measure of entanglement is given by
\begin{equation}
\label{sadher-lau1}
\overline{E}_g(U) = \max_{|\psi\rangle \otimes |\phi\rangle 
} E_k\left(U_{AB}|\psi\rangle_A \otimes |\phi\rangle_B\right).
\end{equation}
Note that $E_1$ is always unity.
As we are concentrating on two-qubit systems, here $k$ runs over 1 and 2. Therefore for our purposes, $k=2$ in Eq.(\ref{sadher-lau1}).

We next move over to quantum coherence generation by the unitary operator, \(U_{AB}\).
To deal with this question, we first need to fix the basis with respect to which the quantum coherence is to be calculated. Let us suppose that this basis of \(\mathcal{H}_A  \otimes \mathcal{H}_B \) is
\begin{equation}
\mathcal{B} = {\ket{\psi_i}}_{i=1}^{d_Ad_B}, 
\end{equation}
where \(d_A = \dim \mathcal{H}_A\) and likewise for $d_B$. 
The  coherence power~\cite{Yao,Winter2}, with respect to a basis $\mathcal{B}$, generated by the unitary, \(U_{AB}\), is given by
\begin{eqnarray}
C_g(U_{AB}|\mathcal{B})= \max_{\varrho_{AB}}\big[ C\left(U_{AB}   \varrho_{AB} U_{AB}^\dagger |\mathcal{B}\right) - C\left( \varrho_{AB} |\mathcal{B}\right)\big],\nonumber\\
\end{eqnarray}
where again, like in the case of entanglement generation, the maximization is over all states, \(\varrho_{AB}\) on \(\mathcal{H}_A \otimes \mathcal{H}_B\). And again, it is possible to consider the generating power by considering an additional optimization over all local extensions on Alice's and Bob's spaces. 
\(C\) is a measure of quantum coherence,
and while several 
quantum coherence measures are known in the literature, 
we consider the relative entropy of quantum coherence and the $l_1$-norm of quantum coherence \cite{Plenio} for our purposes. Just like for the case of entanglement we will consider pure inputs, so that for relative entropy of coherence,
\begin{equation}
\label{coh}
C(|\psi\rangle|\mathcal{B})=S(\rho_{diag}),
\end{equation}
where 
$\rho_{diag}$ is a diagonal density matrix constructed by the diagonal elements of $|\psi\rangle \langle \psi|$, when written in the basis, \(\mathcal{B}\). 
We focus 
on product as well as arbitrary orthonormal bases of the bipartite system, for computing the relative entropy of quantum coherence.
Arbitrary product bases in general bipartite quantum systems is a relatively less understood concept. However, for two-qubit systems, all product orthonormal bases have been characterized~\cite{Walgate}, and they can be expressed as
\begin{equation}
\label{eq:10}
\ket{00},~ \ket{0 1},~ \ket{1 \eta},~ \ket{1 \eta^{\perp}}.
\end{equation}
In the computational basis, the \(|\eta\rangle\) and \(|\eta^\perp\rangle\) can be written as 
\(|\eta\rangle = \cos(\theta/2)|0\rangle + e^{i\phi}\sin(\theta/2)|1\rangle\) and \(|\eta^\perp\rangle = -e^{-i\phi}\sin(\theta/2)|0\rangle + \cos(\theta/2)|1\rangle\) where $0 \le \theta \le \pi$ and $0 \le \phi < 2\pi$.
The elements for an arbitrary two-qubit product basis can be chosen, for our purposes, as in (\ref{eq:10}).
Since we optimize over all input states of two-qubits for a given unitary, we have the freedom of choosing an arbitrary basis as the computational basis on each local qubit space. Therefore, while the general two-qubit product basis is \(|\eta' \eta'' \rangle\), \(|\eta' \eta''^\perp \rangle\), \(|\eta'^\perp \eta \rangle\), \(|\eta'^\perp \eta^\perp \rangle\),
we can use the mentioned freedom to choose \(|\eta'\rangle = |0\rangle\), \( |\eta'^\perp \rangle = |1\rangle\)
for the first qubit and 
\(|\eta''\rangle = |0\rangle\), \( |\eta''^\perp \rangle = |1\rangle\)
for the second qubit.


The construction of an arbitrary two-qubit basis, which would then contain entangled elements, in general, can be attained by applying 
 a two-qubit arbitrary non-local unitary on the four elements of the computational basis, $\{ \ket{00}, \ket{01}, \ket{10}, \ket{11} \}$. The two-qubit unitary can be formed by using the prescription of~\cite{Kraus, Khaneja}, discussed below (see Eq.~(\ref{eq:8})). 
When acted on by the unitary, the computational basis elements, $\ket{00}$, $\ket{01}$, $\ket{10}$, $\ket{11}$, yield four states, which serve as the four elements of an arbitrary basis. 

Just like for the case of entanglement generation, we consider only those inputs for which the resource is vanishing. As we have already mentioned, we only consider pure inputs. Therefore, for quantum coherence generation with respect to a product or arbitrary orthogonal basis, \(\mathcal{B}_{\mathbb{C}^2 \otimes \mathbb{C}^2}\), on the two-qubit Hilbert space, the inputs can only be the four states of \(\mathcal{B}_{\mathbb{C}^2 \otimes \mathbb{C}^2}\). 
So finally, the amount of quantum coherence with respect to the basis, \(\mathcal{B}_{\mathbb{C}^2 \otimes \mathbb{C}^2}\), that is generated by using the unitary, \(U_{AB}\), is given by 
\begin{equation}
\label{lilua-batase}
C_g(U|\mathcal{B}_{\mathbb{C}^2 \otimes \mathbb{C}^2}) = \max_{i} C(U_{AB}|\phi_i\rangle_{AB}|\mathcal{B}_{\mathbb{C}^2 \otimes \mathbb{C}^2}),
\end{equation}
where \(\mathcal{B}_{\mathbb{C}^2 \otimes \mathbb{C}^2} = \{|\phi_i\rangle_{AB}\}_{i=1}^4\). 
Now for obtaining the quantum coherence generating power of the unitary with respect to product bases, we have to take a maximization over  arbitrary product bases of the $\mathbb{C}^2 \otimes \mathbb{C}^2$ Hilbert space. So, we have
\begin{equation}
\label{500miles}
    C_g(U)=\max_{\text{product bases } \mathcal{B}_{\mathbb{C}^2 \otimes \mathbb{C}^2}} \max_{i} C(U_{AB}|\phi_i\rangle_{AB}|\mathcal{B}_{\mathbb{C}^2 \otimes \mathbb{C}^2})
\end{equation}
Similarly, for obtaining the same for arbitrary bases, we have 
\begin{equation}
\label{500miles_ent}
    C_g^{\prime}(U)=\max_{\text{arb bases } \mathcal{B}_{\mathbb{C}^2 \otimes \mathbb{C}^2}} \max_{i} C(U_{AB}|\phi_i\rangle_{AB}|\mathcal{B}_{\mathbb{C}^2 \otimes \mathbb{C}^2}).
\end{equation}
The quantum coherence generating power, with any other measure of quantum coherence, can be defined similarly as the same in Eqs.~(\ref{500miles}) and ~(\ref{500miles_ent}), for product and arbitrary bases respectively, with only the measure $C$ being replaced by some other measure. A measure of quantum coherence that is conceptually different from the relative entropy of quantum coherence is the $l_1$-norm of quantum coherence, being defined, for an arbitrary state $\ket{\psi}$ and a basis $\mathcal{B}$, as
\begin{equation}
    C_{l_1}(|\psi\rangle|\mathcal{B})=\sum_{\substack{i,j\\i \ne j}} |\rho_{i,j}|.
\end{equation}
This is the sum of moduluses of the off-diagonal elements of the density matrix when written in the basis $\mathcal{B}$. Similarly as for the relative entropy of quantum coherence, we can e.g. optimize over all product bases to obtain
\begin{equation}
\label{500miles_l1}
    \overline{C}_g(U)=\max_{\text{product bases } \mathcal{B}_{\mathbb{C}^2 \otimes \mathbb{C}^2}} \max_{i} C_{l_1}(U_{AB}|\phi_i\rangle_{AB}|\mathcal{B}_{\mathbb{C}^2 \otimes \mathbb{C}^2}).
\end{equation}
Henceforth in this paper, we continue with local von Neumann entropy as the entanglement quantifier (for pure bipartite states) as in Eq.(\ref{ent}), and relative entropy of quantum coherence as the quantifier of quantum coherence as in Eq.(\ref{coh}) (for pure states). We will change the quantifier in Sec.~\ref{Sec:5}C.

An 
arbitrary two-qubit unitary operator  can be written in the form \cite{Kraus, Khaneja} 
\begin{equation}
\label{eq:8}
    U_{AB} = U_A \otimes U_B U_d V_A \otimes V_B, 
\end{equation}
where \(U_A, V_A, U_B, V_B \in U(2)\). $U_A$, $U_B$, $V_A$, $V_B$ are 
unitaries acting on the local subsystems, and \(U_d\) is a ``non-local'' unitary on \(\mathbb{C}^2 \otimes \mathbb{C}^2\) having the form,
\begin{equation}
\label{eq:9}
U_d = \exp(-i\alpha_x \sigma_x \otimes \sigma_x -i\alpha_y \sigma_y \otimes \sigma_y -i\alpha_z \sigma_z \otimes \sigma_z), 
\end{equation}
where \(\alpha_x\), \(\alpha_y\), \(\alpha_z\) are real numbers and \(\sigma_x\), \(\sigma_y\), \(\sigma_z\) are the Pauli matrices. 
The relevant ranges for \(\alpha_x\), \(\alpha_y\), \(\alpha_z\) may depend on the resource being generated, as we see below.

For obtaining the maximum entanglement generated by a fixed unitary, we have to perform a maximization over input states. Since we are taking the input states as arbitrary pure product states, we can forget about the $V_A$ and $V_B$, because they will just rotate the space of product states into itself. 
Also, it is of no use, in this case, to apply the $U_A\otimes U_B$, as local unitaries will keep the entanglement unchanged. 
So, while generating entanglement by a two-qubit unitary, it is enough to only deal with $U_d$, as the amount of entanglement generated will not be affected by the  four local unitaries. In Ref.~\cite{Kraus}, it was shown that whenever 
$\alpha_x+\alpha_y \ge \pi/4$ and $\alpha_y+\alpha_z \le \pi/4$, there always exists an input state for which we get maximal entanglement at the output, and outside the region - dictated by the above inequalities - in the \((\alpha_x,\alpha_y, \alpha_z)\)-parameter space, the maximum entanglement is given by 
\begin{equation}
\label{eq:eof}
E = H\left(\frac{1+\sqrt{1-\overline{C}^2}}{2}\right),
\end{equation}
where \(H(\cdot)\) is the binary entropy function, with 
\(\overline{C}\) being given by~\cite{Hill,Wootters}
\begin{equation}
\label{eq:con}
    \overline{C}=\max_{k,l}|\sin(\lambda_k-\lambda_l)|,
\end{equation}
where $k$, $l$ go from 1 to 4 and
\begin{eqnarray}
\lambda_1&=& \phantom{-}\alpha_x-\alpha_y+\alpha_z,\nonumber\\
\lambda_2&=& -\alpha_x+\alpha_y+\alpha_z,\nonumber\\
\lambda_3&=& -\alpha_x-\alpha_y-\alpha_z,\nonumber\\
\label{eq:14}
\lambda_4&=& \phantom{-}\alpha_x+\alpha_y-\alpha_z.
\end{eqnarray}
It is enough to consider the parameter range 
%
%
$\pi/4 \ge \alpha_x \ge \alpha_y \ge \alpha_z \ge 0$, as the maximal entanglement generated is a periodic function in the \((\alpha_x, \alpha_y, \alpha_z)\) parameter space, and completes a full period in this range.


Moving over to the case of maximal quantum coherence generation over product bases and with the unitary \(U_{AB}\), we note that 
%
the product basis \(\{|\phi_i\rangle\}_i\) in Eq.~(\ref{lilua-batase}) is taken to another product basis by \(V_A\otimes V_B\), which will anyway be considered in the maximization over product bases in (\ref{500miles}). This argument is essentially the same for arbitrary bases also. So, in (\ref{500miles}), (\ref{500miles_ent}) and (\ref{500miles_l1}), we can ignore the \(V_A \otimes V_B\). 
\(U_A \otimes U_B\) remains relevant  throughout the quantum coherence part of this paper, and of course \(U_d\) is relevant in both entanglement and quantum coherence parts. 
For the optimizations, 
we have used the 
algorithms of NLOPT~\cite{NLOPT}. 


\section{Quantum coherence generating  power of unitary gate}
\label{Sec:3}
In this section, we try to find the maximum  coherence generating powers of two-qubit unitary operators. 
\begin{figure*}
\includegraphics[height=7cm,width=18cm]{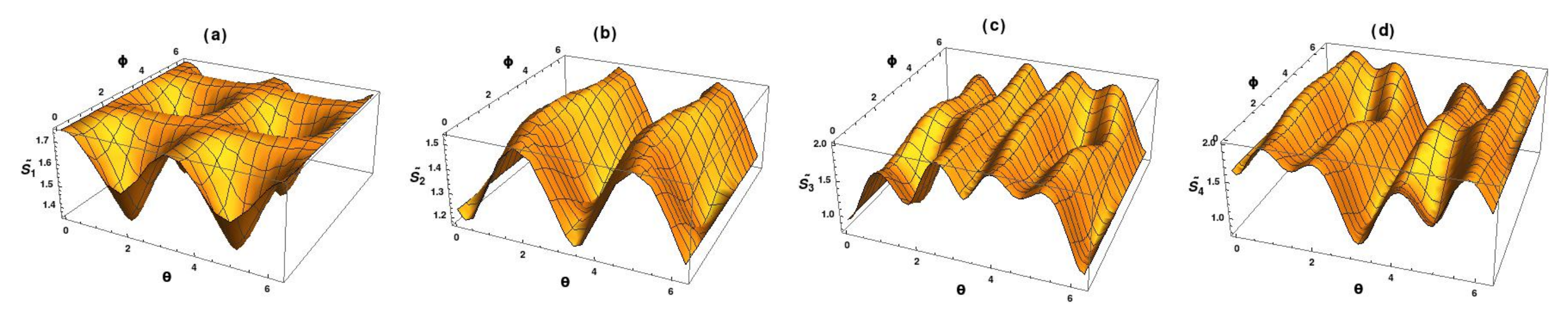}
\caption{Quantum coherence generation for a general two-qubit unitary. We plot here 
the relative entropy of quantum coherence that can be generated by the unitary, $U_{AB}$, as a function of the parameters of the product basis with respect to which the quantum coherence is defined and the elements of which act as initial states for the unitary evolution. The four plots are for the four elements of the product basis that act as the initial states of the evolution. 
The functions $\tilde{S}_1$, $\tilde{S}_2$, $\tilde{S}_3$ and $\tilde{S}_4$ are plotted with respect to the basis parameters, $\theta$ and $\phi$, in panels $(a)$, $(b)$, $(c)$ and $(d)$ respectively. The fixed parameters of the unitary is taken as $\alpha_x=0.6078$, $\alpha_y=0.2625$, $\alpha_z=0.2287$, $\theta_1=2.2330$, $\phi_1=2.1630$, $\psi_1=1.1980$, $\theta_2=3.0700$ $\phi_2=6.0630$, $\psi_2=9.0910$, $\theta_3=1.1000$, $\phi_3=1.6570$, $\psi_3=6.3530$, $\theta_4=1.9190$, $\phi_4=5.2110$ and $\psi_4=11.2600$. 
The quantities $\theta$ and $\phi$ in the horizontal axes are presented in radians and the vertical axes are in bits. 
The \(\theta_i\), \(\phi_i\) and \(\psi_i\), for \(i=1,2,3,4\), are in radians.}
\label{fig5}
\end{figure*}
The unitary is of the form
\begin{equation}
\label{eq:30}
    U_{AB} = U_A \otimes U_B U_d V_A \otimes V_B.
\end{equation}
The \(U_d\) is referred to as the 
``Cartan kernel part'' of the general two-qubit unitary gate. 
The entanglement power of the gate depends only on the Cartan kernel part of the unitary and is not altered due to the presence of the local unitaries. 
However, these local unitaries along with the Cartan kernel part play a role in the quantum coherence power of the unitary. 
We set $W_1=V_A$, $W_2=V_B$, $W_3=U_A$ and $W_4=U_B$, with   $W_k \in \text{SU}(2)$ for $k=1$, 2, 3, 4, represented as
\begin{equation}
\label{unitary}
        W_k=
        \left( {\begin{array}{cc}
        \cos\frac{\theta_k}{2}e^{\frac{i}{2}(\psi_k+\phi_k)} &  \sin\frac{\theta_k}{2}e^{-\frac{i}{2}(\psi_k-\phi_k)} \\
        -\sin\frac{\theta_k}{2}e^{\frac{i}{2}(\psi_k-\phi_k)} &  \cos\frac{\theta_k}{2}e^{-\frac{i}{2}(\psi_k+\phi_k)} 
        \end{array} } \right).
\end{equation}
where $\theta_k \in [0,\pi]$, $\phi_k \in [0,2\pi]$ and $\psi_k \in [0,4\pi)$.
The form of the non-local unitary, $U_d$, remains the same as 
in Eq.~(\ref{eq:9}).
We choose an arbitrary product basis according to Eq.~(\ref{eq:10}), and evaluate the expression for the relative entropy of quantum coherence which is generated by $U_{AB}$ when acting on an incoherent pure two-qubit quantum state. The unitary $U_{AB}$ of the form given in Eq.~(\ref{eq:30}) can be expressed as
\begin{equation}
        U_{AB}=
        \left( {\begin{array}{cccc}
        r_3 & r_1 & m_3 & m_1 \\
        r_7 & r_5 & m_7 & m_5 \\
        r_4 & r_2 & m_4 & m_2 \\
        r_8 & r_6 & m_8 & m_6  
        \end{array} } \right).
\end{equation}
We now calculate the relative entropy of quantum coherence for each of the states in the basis,  
\(\{\ket{00}, \ket{01}, \ket{1\eta}, \ket{1\eta^{\perp}}\}\). 
They are 
respectively given by
\begin{eqnarray}
\label{equ:1}
\tilde{S}_{i} &=& -|r_{5-2i}|^2\log_2|r_{5-2i}|^2-|r_{9-2i}|^2\log_2|r_{9-2i}|^2 \nonumber \\
&-& |r_{10-2i}\cos\frac{\theta}{2}-e^{i\phi}r_{2i-2(-1)^i}\sin\frac{\theta}{2}|^2  \nonumber \\
&& \log_2|r_{10-2i}\cos\frac{\theta}{2}-e^{i\phi}r_{2i-2(-1)^i}\sin\frac{\theta}{2}|^2  \nonumber \\
&-& |r_{2i-2(-1)^i}\cos\frac{\theta}{2}+e^{-i\phi}r_{10-2i}\sin\frac{\theta}{2}|^2  \nonumber \\
&& \log_2|r_{2i-2(-1)^i}\cos\frac{\theta}{2}+e^{-i\phi}r_{10-2i}\sin\frac{\theta}{2}|^2, \nonumber \\
\end{eqnarray}
for $i=1,2$. And for $i=3,4$, the corresponding expressions for quantum coherence are
\begin{widetext}
\begin{eqnarray}
\label{equ2}
\tilde{S}_{i}&=&-|m_{9-2i}\cos\frac{\theta}{2}-(-1)^ie^{-(-1)^i i\phi}m_{2i-5}\sin\frac{\theta}{2}|^2\log_2|m_{9-2i}\cos\frac{\theta}{2}-(-1)^ie^{-(-1)^i i\phi}m_{2i-5}\sin\frac{\theta}{2}|^2 \nonumber \\
&-& |m_{13-2i}\cos\frac{\theta}{2}-(-1)^ie^{-(-1)^i i\phi}m_{2i-1}\sin\frac{\theta}{2}|^2 \log_2 |m_{13-2i}\cos\frac{\theta}{2}-(-1)^ie^{-(-1)^i i\phi}m_{2i-1}\sin\frac{\theta}{2}|^2 \nonumber \\
&-& |\frac{1}{4}\big(m_8(-(-1)^i + \cos\theta) +e^{2i\phi}m_2((-1)^i+\cos\theta) - e^{i\phi}(m_4-m_6)\sin\theta \big)|^2 \nonumber \\
&& \log_2 |\frac{1}{4}\big(m_8(-(-1)^i + \cos\theta) +e^{2i\phi}m_2((-1)^i+\cos\theta) - e^{i\phi}(m_4-m_6)\sin\theta \big)|^2 \nonumber \\
&-& |\frac{1}{4}\big(e^{i\phi}(-(-1)^im_4-(-1)^im_6+(m_4-m_6)\cos\theta)+(e^{2i\phi}m_2+m_8)\sin\theta\big)|^2 \nonumber \\
&& \log_2 |\frac{1}{4}\big(e^{i\phi}(-(-1)^im_4-(-1)^im_6+(m_4-m_6)\cos\theta)+(e^{2i\phi}m_2+m_8)\sin\theta\big)|^2. 
\end{eqnarray}
\end{widetext}
\begin{figure*}
\centering
\includegraphics[width=8cm]{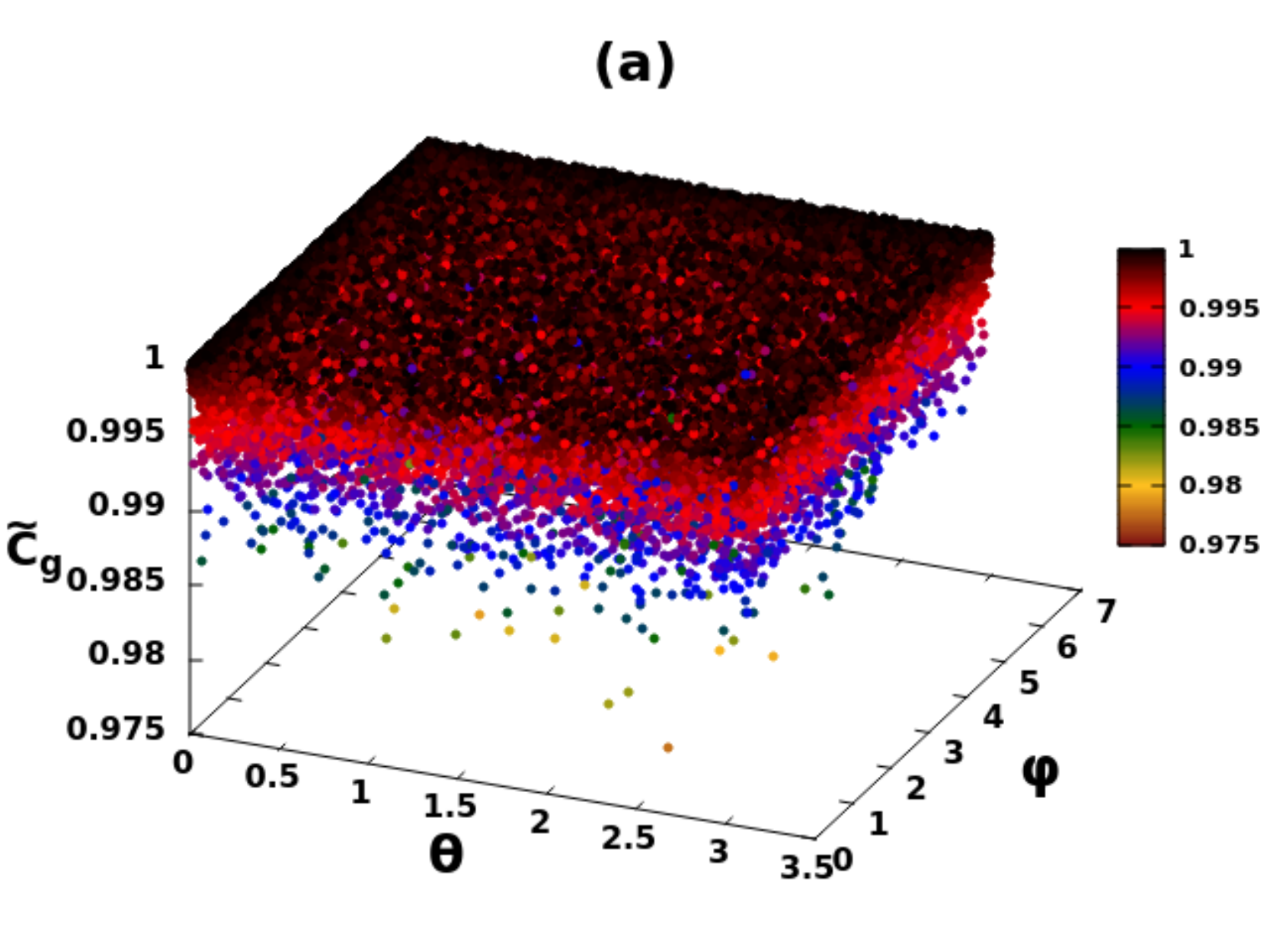}%
\hspace{.25cm}%
\includegraphics[width=8cm]{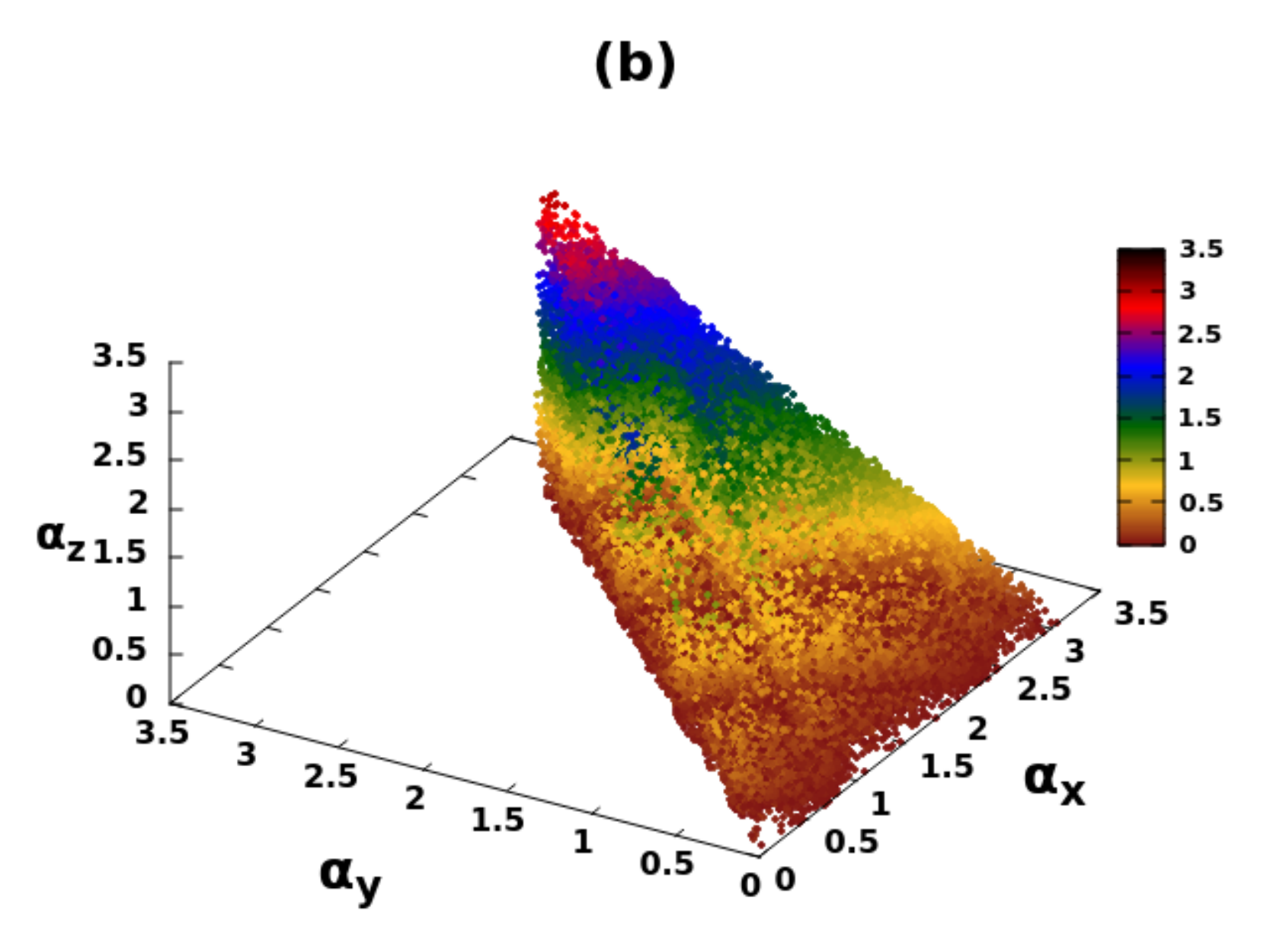}%
\includegraphics[width=8cm]{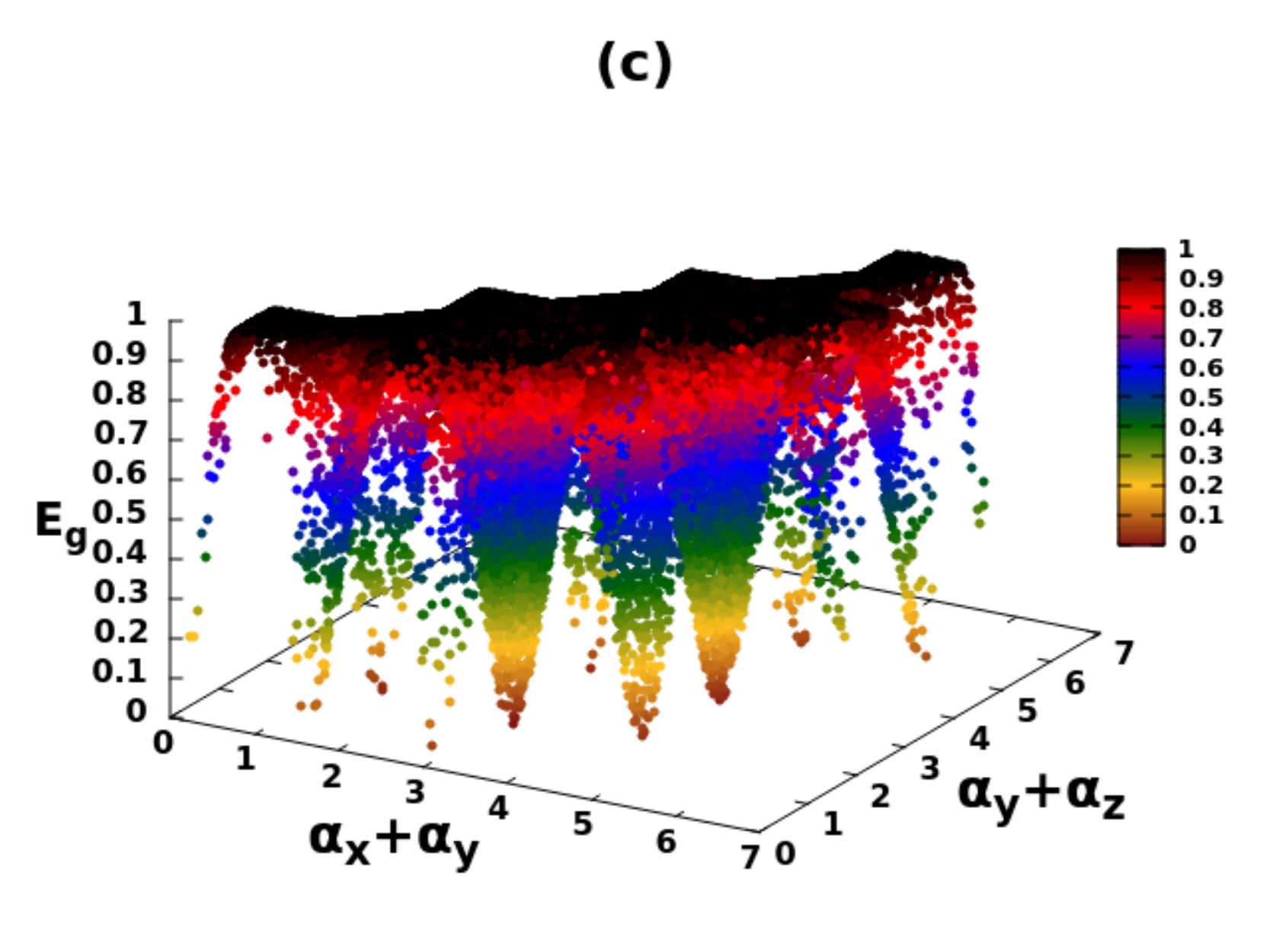}%
\caption{
Highest quantum coherences for maximal entanglement generating general two-qubit unitaries, and vice versa. We investigate here an interplay between two resource generating power of the two-qubit unitaries, \(U_{AB}\). In panel $(a)$, we generate 
2 $\times$ $10^5$ unitaries which can create maximal entanglement, and plot the maximal relative entropy of quantum coherence that can be  generated by the same. The plot is depicted against a base of \(\theta\) along x-axis and \(\phi\) along y-axis, the parameters of the optimal basis for the quantum coherence generation.
In panel \((b)\), 
for 2 $\times$ $10^5$ points,
we mark the region of the parameter space of the general two-qubit unitaries where the generated quantum coherence can reach the maximal value.  
In panel $(c)$, the entanglement power of  unitaries which can generate maximal quantum coherence are presented. For more details, please see the text. The $\theta$, $\phi$ axes are measured in radians. \(\alpha_x\), \(\alpha_y\), \(\alpha_z\) are dimensionless. The quantum coherence used in the plots are measured in bits, the entanglement therein are in ebits.} 
\label{fig9}
\end{figure*}
If we fix the unitary, i.e., if we fix all $r_j$'s and $m_j$'s, we can maximize these four functions with respect to $\theta$ and $\phi$ - the parameters of the product basis. Hence, following Eq.~(\ref{lilua-batase}), the maximum quantum coherence for the fixed unitary and for the fixed basis is $\max\{\tilde{S}_1,\tilde{S}_2,\tilde{S}_3,\tilde{S}_4\}$, where the \(\tilde{S}_i\) for $i=1,2$ are given 
in Eq.~(\ref{equ:1}) and for $i=3,4$ are in~(\ref{equ2}).
So, in accordance with Eq.~(\ref{500miles}), the maximum relative entropy of quantum  coherence generated by the fixed unitary, \(U_{AB}\), is given by 
\begin{equation}
\label{eq:21}
    C_{g}(U_{AB})=\max_{\theta,\phi}[\max\{\tilde{S}_1,\tilde{S}_2,\tilde{S}_3,\tilde{S}_4\}].
\end{equation}
All the four functions $\tilde{S}_i$
are periodic with respect to each of $\alpha_x$, $\alpha_y$ and $\alpha_z$ with a period of $\pi$. So, for evaluating the generation of relative entropy of quantum coherence, we can use the bounds, $\pi \ge \alpha_x,\alpha_y,\alpha_z \ge 0$, because beyond this region, the functions repeat their natures.
In Fig.~\ref{fig5}, we  plot the four 
$\tilde{S}_i$'s with respect to the basis parameters for a fixed unitary. The numerical values of the  parameters of the unitary considered here are given in the caption of Fig.~\ref{fig5}. 
There are several maxima
with respect to $\theta$ in $[0,\pi]$. 
The last two functions, $\tilde{S}_3$ and $\tilde{S}_4$, are neither periodic in $\theta \in [0,\pi]$ nor symmetric around $\theta=\pi$. (The functions are of course periodic in $\theta$ with a periodicity $2\pi$.) On the contrary, $\tilde{S}_1$ and $\tilde{S}_2$ are periodic in $\theta \in [0,\pi]$.
It can be cumbersome to obtain the best choice of basis analytically, but it can be done numerically, 
using globally convergent routines.
For a fixed unitary, we have observed numerically that the maximum quantum coherence corresponding to the initial states $\ket{0 0}$ and $\ket{0 1}$ (the maximum values of $\tilde{S}_1$ and $\tilde{S}_2$) are different. But, for 
the states $\ket{1 \eta}$ and $\ket{1 \eta^{\perp}}$, the maximum values of $\tilde{S}_3$ and $\tilde{S}_4$ are equal. 
Here, along with  possibility that the $\tilde{\theta}_{m_i}$ and $\tilde{\phi}_{m_i}$'s are different for different \(i\), each of the functions, $\tilde{S}_i$, can have multiple maxima in the parameter space of $\theta$ in $[0,\pi]$ and $\phi$ in $[0,2\pi)$. So, we have that for a general unitary, the best choice of basis for quantum coherence generation is not unique and the number of best bases can even be more than four. We have numerically observed that 
the four functions, $\tilde{S}_{i}$, are periodic in each of $\alpha_x$, $\alpha_y$ and $\alpha_z$, and considering the 
parameter range $[0,\pi]$ for the $\alpha$'s is sufficient, and we use this information for further analysis and illustration of the quantum coherence power of $U_{AB}$'s. 
\subsection{Dependence of resource generating power on  parameters of $U_{AB}$}

We try to compare here between the 
two resource generating powers (entanglement and quantum coherence) 
of the general unitary, $U_{AB}$.
As mentioned earlier, the local unitaries $V_A$ and $V_B$ have no contribution in the maximal creation of entanglement and they also do not affect the maximum generated quantum coherence.
So, here we do the analysis and illustration by discarding the $V_A \otimes V_B$ part and considering $\tilde{U}_{AB}=U_{A} \otimes U_B U_d$ as equivalent to the whole unitary $U_{AB}$. 
In an actual experimental implementation, this can necessitate a local rotation of the optimal input state.

\paragraph{Highest quantum coherence generated by \(U_{AB}\) that allows maximal entanglement generation.}
We now find the maximum quantum coherence that can be generated 
by the unitary operators, $\tilde{U}_{AB}$, which can maximize the entanglement  of a bipartite quantum state to the maximal value, i.e., by those \(\tilde{U}_{AB}\) for which $E_g(\tilde{U}_{AB}) =1$. Kraus and Cirac~\cite{Kraus} have shown that the maximal entanglement generated by a two-qubit unitary has certain periodicity and symmetry properties so that it is enough to consider the range $\pi/4 \ge \alpha_x \ge \alpha_y \ge \alpha_z \ge 0$. As previously discussed, the parameter region is bounded by 0 to $\pi$ in case of evaluating the maximal quantum coherence generated, and this is the range that we use therefore for numerically or analytically analyzing quantum coherence generation or its interplay with entanglement generation.
For this analysis, we first evaluate $\tilde{U}_{AB}\ket{\psi}_A\otimes \ket{\phi}_B $ for arbitrarily chosen parameters of the unitary,  
and then calculate the local von Neumann entropy of the output state and optimize over all $\ket{\psi}_A, \ket{\phi}_B \in \mathbb{C}^2$, where $\ket{\psi}_A$ and  $\ket{\phi}_B$ are taken as
\begin{eqnarray}
\label{eq:36}
\ket{\psi_A} &=&  \cos(\Bar{\alpha}/2)\ket{0_A} + e^{i\Bar{\beta}}\sin(\Bar{\alpha}/2) \ket{1_A},\nonumber\\
\ket{\phi_B} &=& \cos(\Bar{\gamma}/2)\ket{0_B} + e^{i\Bar{\delta}}\sin(\Bar{\gamma}/2) \ket{1_B}.
\end{eqnarray}
Therefore, to find the best entanglement generation, the maximization will be over $\Bar{\alpha},\Bar{\gamma}  \in [0,\pi]$ and $\Bar{\beta},\Bar{\delta}  \in [0,2 \pi)$, of $S(\text{tr}_{A/B} P[\tilde{U}_{AB} \ket{\psi}_A \otimes \ket{\phi}_B])$, for every set of values of the parameters 
$\tilde{U}_{AB}$, which are
$\alpha_x$, $\alpha_y$, $\alpha_z$, $\theta_i$, $\phi_i$ and $\psi_i$, where $i$ stands for 3 and 4. 
%
%
%
We choose $2 \times 10^5$ unitaries in this parameter space, which can create maximal entanglement, and search for the maximum relative entropy of quantum coherence maximized over arbitrary pure product bases.  
The algorithm Direct-L~\cite{Gablonsky} of NLOPT is used for the optimization. 
We get a finite probability of finding the maximal quantum coherence. See Fig.~\ref{fig9}$(a)$. The maximum value of relative entropy of quantum coherence for a two-qubit system 
is $2$ bits, but here, for easy comparison with the maximal value of generated entanglement, we normalize the generated quantum coherence as \begin{equation}
\label{debesh-ray}
\tilde{C}_g(\tilde{U}_{AB})=(1/2)C_g(\tilde{U}_{AB}),
\end{equation}
and so we get the maximum quantum coherence equal to 1 bit instead of 2 bits. 
We remember that the maximal value of generated entanglement in a two-qubit system, according to the definition that we have used, is 1 ebit.
It may be noted that instead of numerically optimizing entanglement with respect to the parameters of $\tilde{U}_{AB}$ 
to identify the unitaries which can generate maximal entanglement, one can also 
decompose any arbitrary two-qubit unitary matrix, $U_{AB}$, into the form given in Eq.~(\ref{eq:30}) following the procedure suggested in~\cite{Kraus},
and then the constraints on the parameters of the unitaries for generating  $E_g(U_{AB})=1$ can be used. 
However, to keep using the expressions for entanglement generation from Ref.~\cite{Kraus}, we must identify the rule for going from the bigger range of parameters to the smaller one. This is as follows.

We choose the \(\alpha_i\), where $i$ can be $x$ or $y$ or $z$, in the range \([0,\pi]\), and if we wish to find the maximal entanglement generated by the corresponding unitary,  
we set \(\tilde{\alpha}_i = \alpha_i\) if \(\alpha_i \leq \pi/4\) and \(=\pi/2 - \alpha_i\) if \(\pi/4 \leq \alpha_i \leq \pi/2\), and using the symmetry of generated entanglement around \(\pi/4\), \(U_d(\{\tilde{\alpha}_i\})\) and \(U_d(\{\alpha_i\})\) generates the same entanglement. As previously mentioned, the local unitaries $U_A$, $U_B$, $V_A$ and $V_B$ can be ignored in the case of entanglement generation. If \(\alpha_i \in [\pi/2,\pi]\), we first use periodicity of the generated entanglement and go to an equivalent set of parameters \(\alpha_i^\prime\) - equivalent with respect to entanglement generation - where \(\alpha_i^\prime = \alpha_i -\pi/2\). If \(\alpha_i^\prime \in [0,\pi/4]\), then we set \(\tilde{\alpha}_i = \alpha_i^\prime\), and if \(\alpha_i^\prime \in [\pi/4,\pi/2]\), then we use the symmetry of the entanglement generated and set \(\tilde{\alpha}_i = \pi/2-\alpha_i^\prime\).

So, now we have all the $\tilde{\alpha}_i$ in the parameter range 0 to $\pi/4$, but the order, $\alpha_x \ge \alpha_y \ge \alpha_z$, that was present in the range 0 to $\pi$, may get altered in case of the $\tilde{\alpha}_i$'s, as the symmetry about $\pi/4$ of generated entanglement is a reflection symmetry. Hence, the conditions $\alpha_x+\alpha_y \ge \pi/4$ and $\alpha_y+\alpha_z \le \pi/4$ for obtaining maximal entanglement in case when the \(\alpha_i\) belong to the range \([0,\pi/4]\),  have to be modified. Suppose that after the mapping, we have the order, $\tilde{\alpha}_i \ge \tilde{\alpha}_j \ge \tilde{\alpha}_k$, where $i$, $j$, $k$ are from the set,   \(\{x,y,z\}\), and there are no repetitions. The conditions for getting maximal entanglement will then be $\tilde{\alpha}_i+\tilde{\alpha}_j \ge \pi/4$ and $\tilde{\alpha}_j+\tilde{\alpha}_k \le \pi/4$. Therefore, we  have the $\alpha_i$'s in the parameter space 0 to $\pi$ which can generate maximal entanglement. And then we can, we wish to analyze the maximal quantum  coherence that can be generated by unitaries that can generate maximally entangled states.

\paragraph{Maximal quantum coherence generating unitaries.} 
We now try to identify the $\alpha_x$, $\alpha_y$ and $\alpha_z$, corresponding to which we get $\tilde{C}_g(\tilde{U}_{AB})=1$, when we act the unitary operator $\tilde{U}_{AB}$ on an incoherent two-qubit pure quantum state.
In Fig.~\ref{fig9}$(b)$, 
we depict the region of the \((\alpha_x,\alpha_y, \alpha_z)\)-space  for random choice of the remaining parameters (i.e., 
$\theta_j$, $\phi_j$ and $\psi_j$, for $j$ running over  3 and 4), at which the unitary, \(\tilde{U}_{AB}\), can lead to maximal quantum coherence (maximization using Isres~\cite{Runarsson}). 



\paragraph{Highest entanglement generated by \(U_{AB}\) that allows maximal quantum coherence generation.} 
\label{beech-bhamar}

In Fig.~\ref{fig9}$(c)$, we look at the entanglement that can be generated by the unitaries which can generate maximum quantum coherence while operating on the set of two-qubit incoherent pure states. 
The unitaries are chosen from Fig.~\ref{fig9}\((b)\).
We use the parameters \(\alpha_x+\alpha_y\) and \(\alpha_y+\alpha_z\) as axes of the base against which the generated entanglement is depicted. Note however that these \(\alpha_x\), \(\alpha_y\), \(\alpha_z\) belong to the range \([0,\pi]\), and so for calculation of the maximal entanglement generation, we need to go to the range \([0,\pi/4]\). And since there is a reflection symmetry that is used in the transformation between the two ranges, the ordering between the \(\alpha\)'s is lost. Consequently, the condition for reaching maximal entanglement from Ref.~\cite{Kraus} cannot be used directly in the presentation in panel~\((c)\). In Fig.~\ref{fig9}$(c)$, we find a finite probability of points (representing unitaries) for which the generated entanglement is 1 ebit. 
From panels $(a)$ and $(c)$ in the same figure, we can conclude that there exists unitaries for which both entanglement and quantum coherence generations are maximal. 
\section{Resource generating  powers of  paradigmatic quantum gates}
\label{Sec:4}

We choose here a few paradigmatic two-qubit gates that are widely used in quantum device circuits, and compare their entanglement and quantum coherence power. 
For completeness, we first define the gates, and then present a table containing the capacities. 

Arguably, the most well-known two-qubit gate is the controlled-NOT (CNOT) operator. The NOT gate is the same as the Pauli-\(\sigma_x\) operator.
The CNOT operator is defined on the two-qubit space as 
\(U_{CNOT} = |0\rangle\langle 0| \otimes I_2 + |1\rangle \langle 1| \otimes \sigma_x\), where \(I_2\) is the identity operator on the qubit space.
The CNOT  is therefore the controlled-\(\sigma_x\) operator. Similarly, one can consider the controlled-\(\sigma_z\) operator as
\(U_{CZ} = |0\rangle\langle 0| \otimes I_2 + |1\rangle \langle 1| \otimes \sigma_z\), which is usually referred to as the CZ gate. The SWAP  is a two-qubit linear operator defined as 
\(U_{SWAP}|\psi\rangle|\phi\rangle = |\phi\rangle|\psi\rangle\), where \(|\psi\rangle,|\phi\rangle \in \mathbb{C}^2\).
[One can of course define a swap gate in arbitrary bipartite dimensions, \(\mathbb{C}^d \otimes \mathbb{C}^d\).]
The SWAP of course cannot create any entanglement by acting on a product state. The situation is very different for its square root, for which the matrix representation in the computational basis, is
%
\begin{equation}
        U_{\sqrt{SWAP}}=
        \left( {\begin{array}{cccc}
        1 & 0 & 0 & 0 \\
        0 & \frac{1}{2}(1+i) & \frac{1}{2}(1-i) & 0 \\
        0 & \frac{1}{2}(1-i) & \frac{1}{2}(1+i) & 0 \\
        0 & 0 & 0 & 1  
        \end{array} } \right).
\end{equation}

So far, we have not considered any product unitary (i.e., product of two single-qubit unitaries), which trivially have vanishing entanglement power. They however can have nontrivial quantum coherence power. We consider three such product unitary gates. The first one that we choose is \(U_{YX} = \sigma_y \otimes \sigma_x\). One can of course consider any other combination of the Pauli spin-1/2 operators. An important single-qubit unitary, that is almost universally present in a quantum algorithm circuit, is the Hadamard gate, defined by 
\(H|0\rangle = \frac{1}{\sqrt{2}}(|0\rangle + |1\rangle)\), 
\(H|1\rangle = \frac{1}{\sqrt{2}}(|0\rangle - |1\rangle)\), supplemented by linearity. 
The other two product unitaries that we consider are \(H\otimes H\) and \(H\otimes I_2\).


In the following table, we present the capacities of entanglement and quantum coherence generation for the above unitaries, correct to four significant figures. 

\begin{center}
\begin{tabular}{ | m{3cm} | m{2.5cm}| m{2cm} | }   
\hline
    \bf{Gate}~$(U)$  & \bf{} $E_g(U)$  & \bf{ $\tilde{C}_g(U)$} \\ 
  \hline
  \hline
  CNOT & 1 & 0.5 \\ 
  \hline
  CZ & 1 & 0.5 \\ 
  \hline
  SWAP & 0 & 0.7768 \\ 
  \hline
  $\sqrt{\text{SWAP}}$ & 1 & 0.75 \\ 
  \hline
  YX & 0 & 0.5 \\
  \hline
  $H \otimes H$ & 0 & 1 \\
  \hline
  $H \otimes  I_2$ & 0 & 0.75 \\
  \hline
\end{tabular}
\end{center}

In the succeeding section, we look at Haar uniformly generated random two-qubit unitaries, and compare their entanglement and quantum coherence powers.


\begin{figure}
\includegraphics[width=8cm]{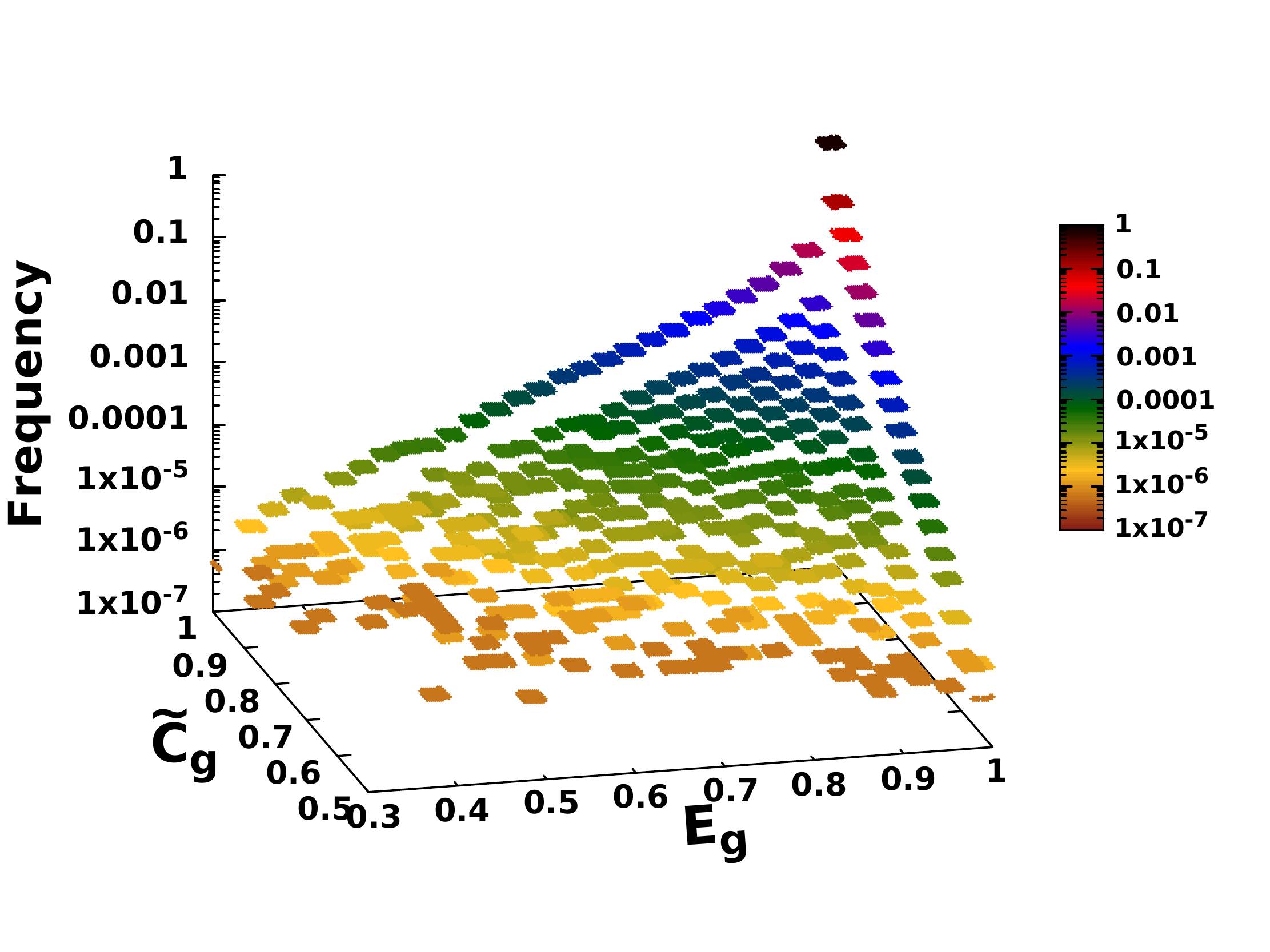}
%
\caption{Resource generating tendencies  of Haar random two-qubit unitaries. 
We Haar uniformly generate a large number of two-qubit unitaries, and find their entanglement and quantum coherence power.
We then divide the \((E_g,\tilde{C}_g)\) space into small squares, and find the relative frequencies of  the number of unitaries that fall in those squares.  
These relative frequencies are plotted along the vertical axis, which is on a logarithmic scale. 
%
The quantity plotted in the vertical axis is dimensionless, while the \(E_g\) and \(\tilde{C}_g\) are in ebits and bits respectively. Please see text for more details.
}
\label{fig6}
\end{figure}

\section{Resource generating  powers of Haar random quantum gates}
\label{Sec:5}
\begin{figure*}
\centering
\includegraphics[width=8cm]{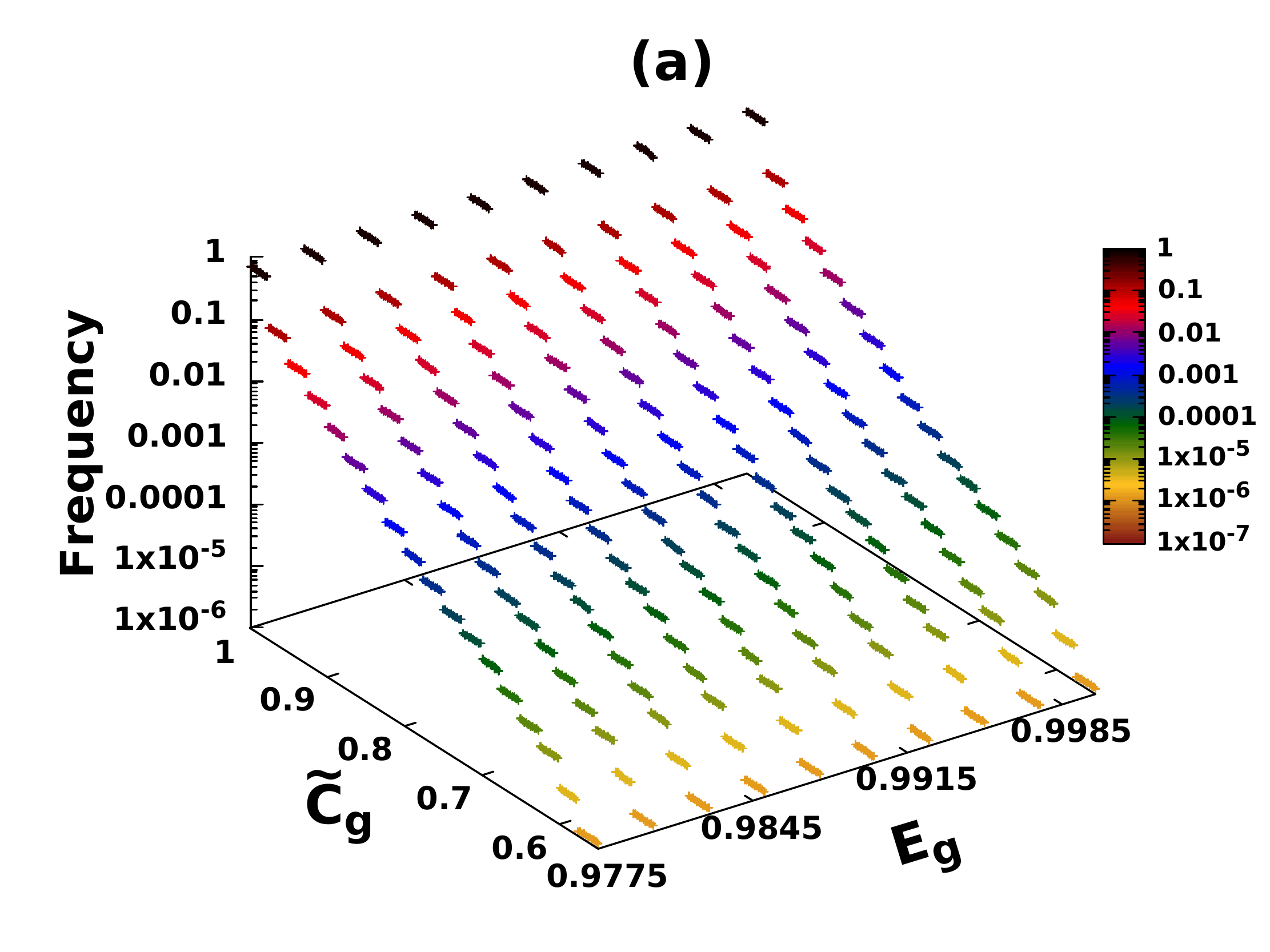}%
\hspace{.25cm}%
\includegraphics[width=8cm]{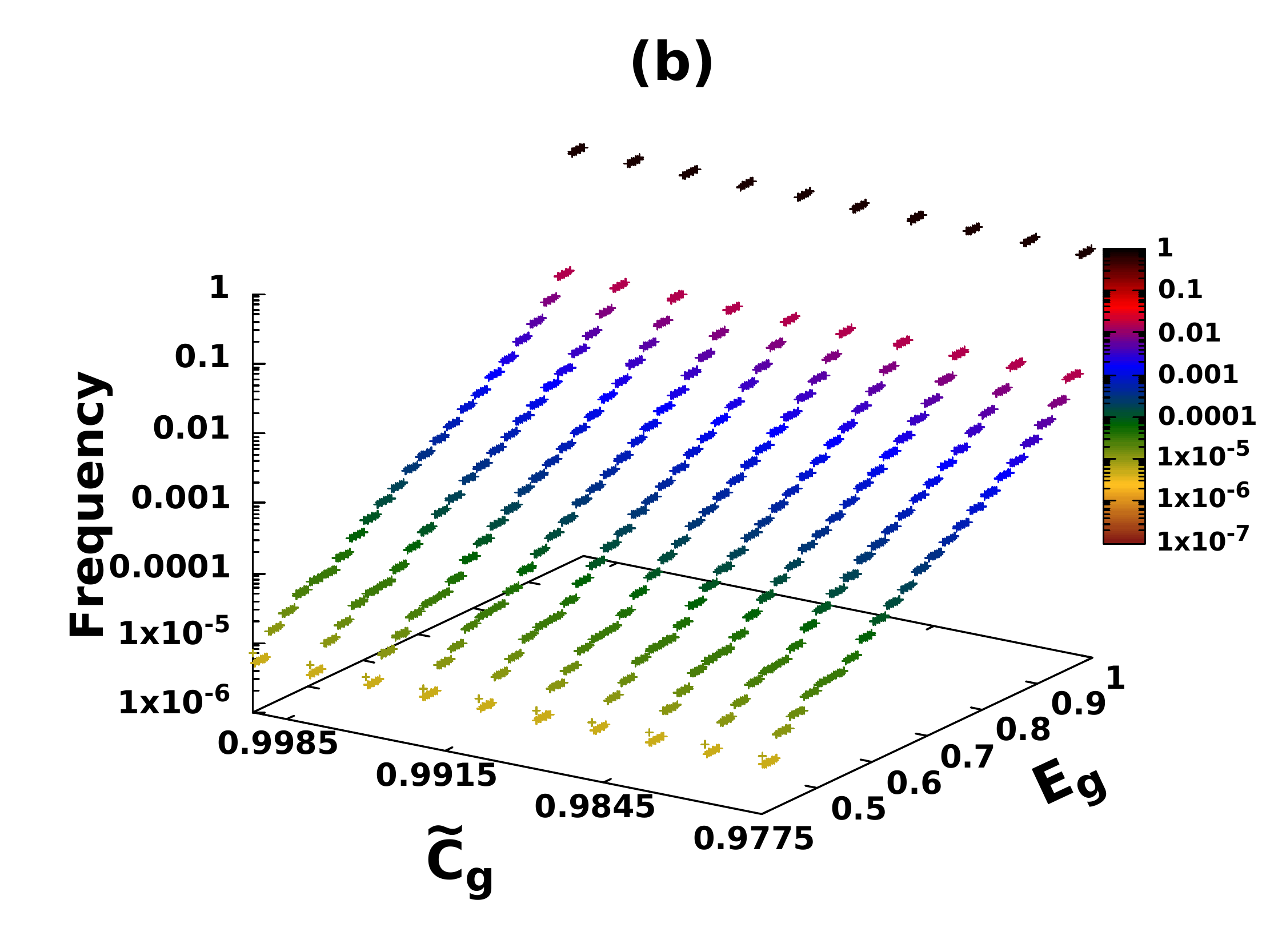}%
\hspace{.25cm}%
\includegraphics[width=8cm]{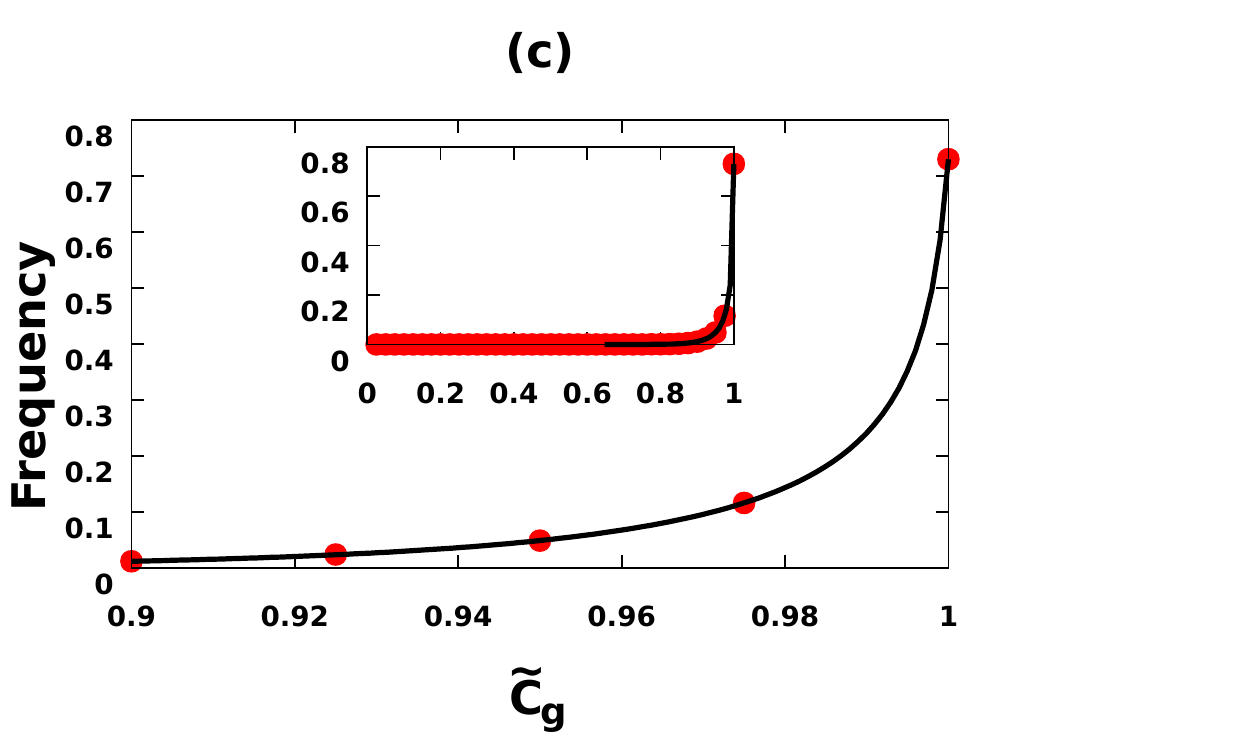}%
\hspace{.25cm}%
\includegraphics[width=8cm]{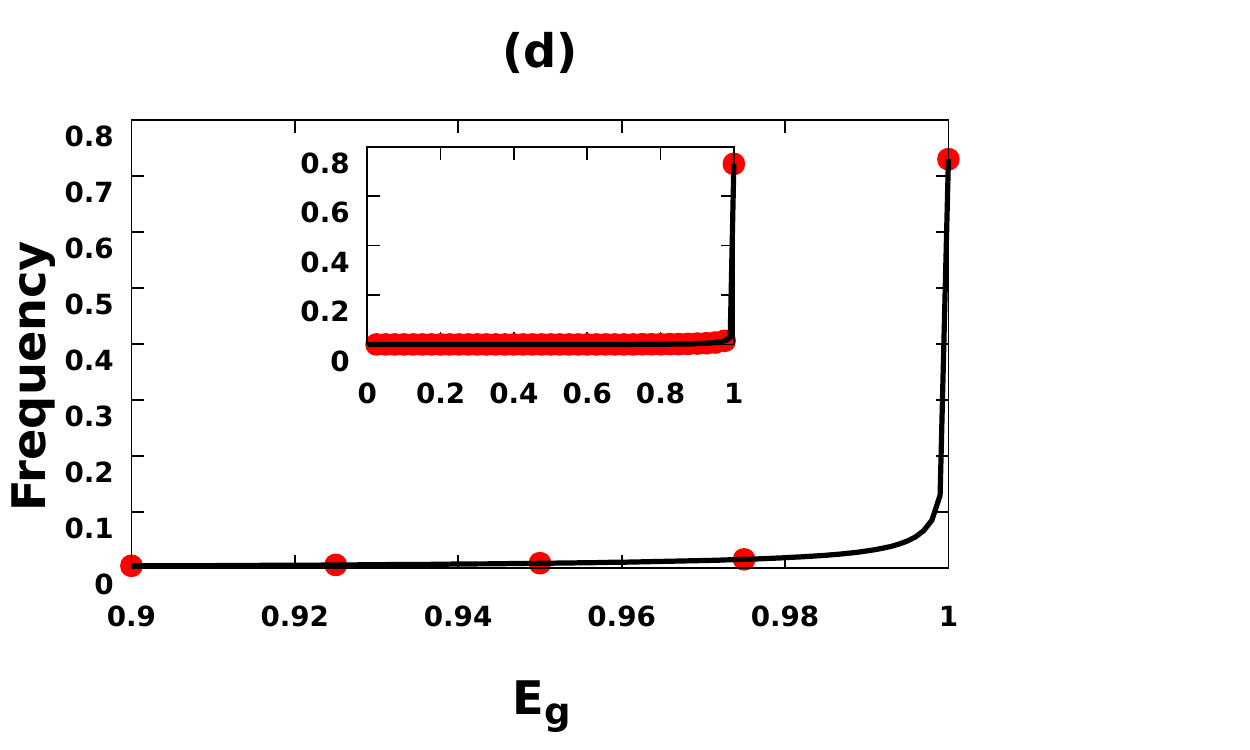}%
\caption{
The asymmetrical nature of resource generating powers of two-qubit unitaries. 
In panel $(a)$, we present a magnified view of the plot in Fig.~\ref{fig6} when restricted to the range \([0.9775,1]\) on the entanglement axis. In panel \((b)\), the same is done for quantum coherence. 
The vertical axes are again in logarithmic scale for both the panels, \((a)\) and \((b)\). 
Just one cross-section of the surface in panel \((a)\) is analyzed in panel \((c)\). The chosen cross-section is for 
$E_g=1$. The inset and the main plot in panel \((c)\) differ only in the range of the horizontal axis. Please also note that the vertical axis is in the normal scale in panel \((c)\). The analysis in panel \((d)\) is exactly the same as in panel \((c)\), with the roles of entanglement and quantum coherence being reversed.
The quantity plotted in the vertical axes is dimensionless, while on the horizontal axes, \(E_g\) is in ebits and \(\tilde{C}_g\) is in bits. Please see text for more details.
}
\label{fig7}
\end{figure*}
In this section, we will compare the entanglement and quantum coherence generating powers of Haar uniformly generated gates on  two-qubit systems.
%
The Haar uniform generation is effected by using the 
Ginibre ensemble~\cite{Ginibre}. For each unitary, 
we numerically evaluate the \(E_g\), by optimizing over 
arbitrary two-qubit pure product states as inputs (Eq.~(\ref{eq:36})). 
For the same unitary, we also obtain $\tilde{C}_g$ 
for a pure incoherent input state, and optimize over the inputs as well as the product (Eq. (\ref{eq:10})) or arbitrary bases, with respect to which the quantum coherence is defined. 
The two maximizations are done independently, so that the input state for which the maximum entanglement is attained, can be different from the state for which we obtain the maximum quantum  coherence. 
Instead of numerically maximizing the entanglement for a unitary, one can also use the canonical decomposition given in~\cite{Kraus}. We present our observations on the resource generating powers of Haar uniformly generated unitaries in the next three subsections. The first two contains the discussions using the measures of entanglement and coherence as the local von Neumann entropy and the relative entropy of coherence respectively. In the third one, we illustrate the results of our investigations, considering the Nielsen-Vidal monotone and $l_1$-norm of coherence, defined in Eqs.~(\ref{sadher-lau1}) and~(\ref{500miles_l1}) respectively, as measures of the corresponding resources.
\subsection{Maximal entanglement vs maximal quantum coherence in product bases}
In Fig.~\ref{fig6}, we depict $E_g$ and 
$\tilde{C}_g$ on the base axes, and $\nu$, the relative frequency of the number of unitaries generating the corresponding $E_g$ and $\tilde{C}_g$, plotted along the vertical axis.
Of course, a finite precision is needed for calculating the relative frequencies, and 
we have a total of $1.6 \times 10^{3}$ squares, with each being of area $2.5 \times 10^{-2}\; \text{ebits} \times 2.5 \times 10^{-2}\; \text{bits}$ on the entanglement-quantum coherence plane. We generate a  total of $1.6 \times 10^6$ two-qubit unitaries, Haar uniformly, and for each square, we plot (on the vertical axis) the relative frequency of the number of unitaries having the ability of generating the numerical values of resources in ebits and bits corresponding to that square.



This depiction describes how the relative frequency  increases progressively from $\approx 0$ to $\approx 1$, as we go along the plane in the direction of increasing entanglement and increasing quantum coherence, and reaches a peak value at the point where entanglement and quantum coherence are both maximal.
Note that the vertical axis has a logarithmic scale in the depiction, and this implies that there is a large fraction of unitaries, in the space of two-qubit unitaries, for which both entanglement and quantum coherence generation are near maximal.


In Fig.~\ref{fig7}$(a)$, we have focused attention at high-entanglement end of Fig.~\ref{fig6}. Precisely, we have restricted 
the entanglement range to \([0.9775,1]\), in ebits. Similarly, Fig.~\ref{fig7}$(b)$ depicts the same in the high-quantum coherence range, \([0.9775,1]\), in bits. We find that there is asymmetry between the entanglement and quantum coherence generations, near their respective maximal values. 


In panels $(c)$ and $(d)$ of Fig.~\ref{fig7} we look at both the resources by first fixing the other resource at a fixed value. Precisely, we take the fixed values to be the maximal ones in both cases. 
%
In panel $(c)$, the fixed resource is entanglement, while in panel $(d)$, it is quantum coherence. The asymmetric nature between entanglement and quantum coherence generation, as seen in the relative frequencies of Haar random unitaries, that we had mentioned before, is now visible more clearly.
We find that the numerically generated red dots in the panels \((c)\) and \((d)\) can be well-described by the beta distribution, suitably scaled and shifted. But the parameters of the beta distributions that fit the two cases are different. 
%
The fitting function therefore has the form,
\begin{equation}
\label{beta}
f_B(x) = d B(x; \alpha_B, \beta_B) + h,
\end{equation}
for \(x\in [0,1]\), and \(\alpha_B>0\), \(\beta_B>0\). The explicit form of the beta distribution, \(B(x; \alpha_B, \beta_B)\), is given in Appendix~\ref{chnader-chhaya}. 
 For Fig.~\ref{fig7}$(c)$, the best fit values of the exponents of $f_B(x)$ are as follows:
\begin{eqnarray}
&&\alpha_B=21.0539 \quad  (\pm 1.244), \quad \nonumber \\
 &&\beta_B=0.4694 \quad (\pm 3.057), \quad \nonumber \\
 &&\text{Error} = 1.85 \times 10^{-4}. 
 \end{eqnarray}
 The numbers in brackets indicate the respective 95\% confidence intervals, and the ``Error" mentioned is the minimum \(\chi^2\) error.
The same numbers
in Fig.~\ref{fig7}$(d)$ are
 \begin{eqnarray}
 &&\alpha_B=9.1513 \quad  (\pm 0.1627), \quad \nonumber \\
 &&\beta_B=0.3766 \quad (\pm 0.0097), \quad \nonumber \\
 &&\text{Error} = 1.186 \times 10^{-3}. 
 \end{eqnarray}
 See Appendix~\ref{chnader-chhaya} for the values of the other parameters.
%
We have used  non-linear least-square fitting to obtain the values of the parameters, their 95\% confidence intervals, and the error estimates, for the fitting curves~\cite{banalata}. 

%
%
\begin{figure}
\includegraphics[width=8cm]{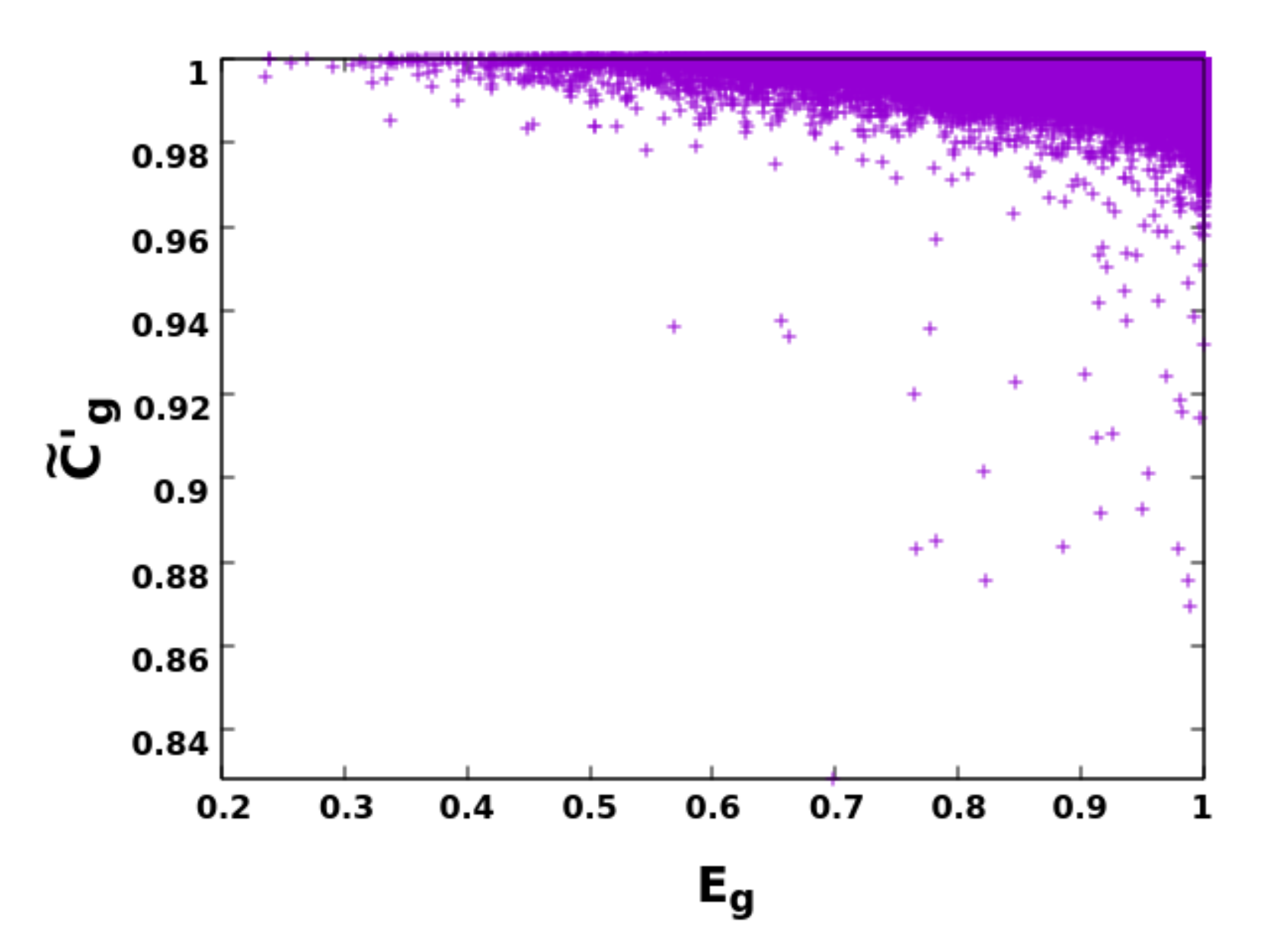}
\caption{Resource generating power of Haar random two-qubit unitaries. Here we demonstrate a scattered plot of the maximum coherence $\tilde{C}_g^{\prime}$ vs maximum entanglement $E_g$ by generating $16 \times 10^{5}$ two-qubit unitaries Haar uniformly. 
Among the quantites plotted here, $E_g$ is in ebits and $\tilde{C}_{g}^{\prime}$ is in bits.}
\label{ent_bas}
\end{figure}
\begin{figure*}
\includegraphics[width=8cm]{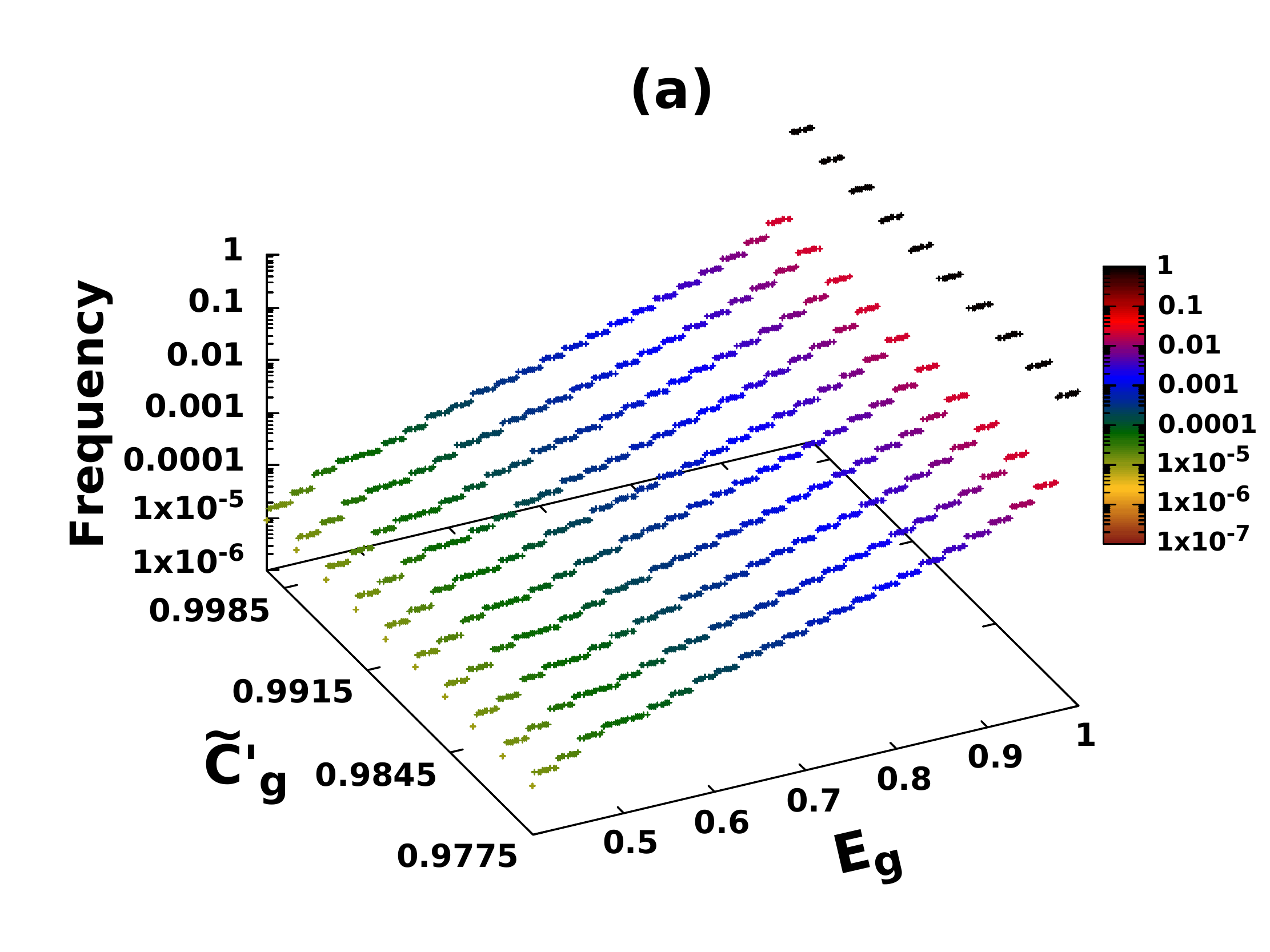}%
\hspace{.40cm}%
\includegraphics[width=8cm]{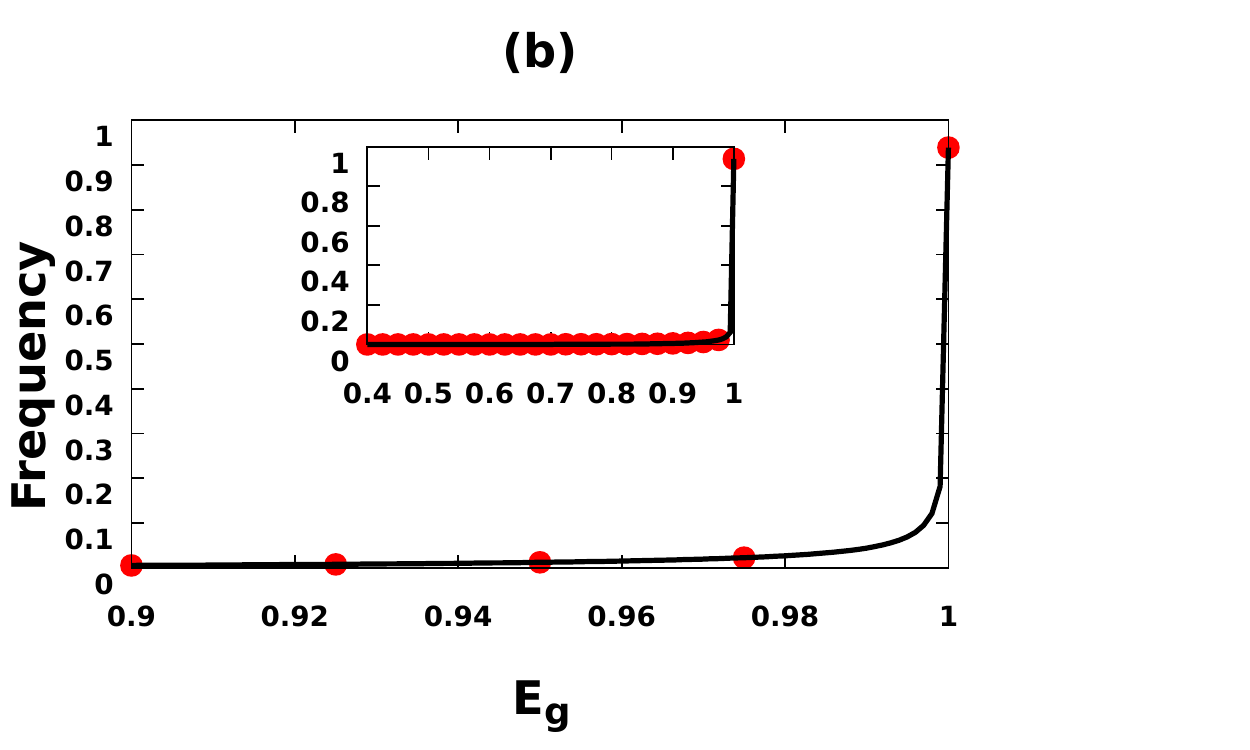}
\caption{Relative frequency vs the two resources generated by the Haar uniform random unitaries, where the optimum coherence is obtained by maximizing over all  bases. In panel (a) the relative frequency is plotted in the range [0.9975,1] on the coherence axis. 
One cross section of (a) for $\tilde{C}_g^{\prime}=1$ is presented in panel (b). All other considerations are same as in Fig.~\ref{fig7}.
The quantities, $E_g$ and $\tilde{C}_{g}^{\prime}$, are expressed in ebits and bits respectively on the horizontal axes, and the quantity represented on the vertical axis is dimensionless.}
\label{ent_bas_freq}
\end{figure*}
\subsection{Maximal entanglement vs maximal quantum coherence in arbitrary bases}
Here we depict $\tilde{C}_g^{\prime}$ with $E_g$ as a scattered plot, in Fig.~\ref{ent_bas}. $\tilde{C}_g^{\prime}$ is just the normalized version of $C_g^{\prime}$ : $\tilde{C}_g^{\prime}(U_{AB})=(1/2)C_g^{\prime}(U_{AB})$.
The choice of zero resource inputs and the methods of optimization are the same as in the preceding case of arbitrary product bases. 
Just like the preceding case, here also we observe the tendency of unitary gates to produce maximal entanglement along with maximal coherence. Most of the points of the scattered plot, depicted for $16 \times 10^{5}$ Haar uniformly generated two-qubit unitaries, are concentrated near the region having both coherence and entanglement close to unity with a higher spread along the entanglement axis. We can observe that the majority of unitaries produce $\tilde{C}_{g}^{\prime}$ approximately $\ge 0.96$ and there exist a finite probability to generate $\tilde{C}_{g}^{\prime}$ from $0.84$ to $0.96$. Typically, there appears almost no possibility to generate $\tilde{C}_{g}^{\prime}< 0.84$.

We now depict the relative frequency of the number of unitaries generating $E_g$ and $\tilde{C}_g^{\prime}$, with the corresponding $E_g$ and 
$\tilde{C}_g^{\prime}$ along the base axes in Fig.~\ref{ent_bas_freq}(a), in a manner similar to that in Fig.~\ref{fig7}(b). We can see that the nature of the frequency in the former one is qualitatively similar to that in the latter one, with a progressively increasing frequency from $\approx 0$ to $1$, reaching the maximum at $E_g=1$.
In panel (b) of Fig.\ref{ent_bas_freq}, we plot a cross-section of panel (a) at $C_g^{\prime}=1$,   and in this case also, the relative frequency exhibits a nature which can be fitted with the beta function given in Eq.~(\ref{beta}). 
The best fit values for the parameters of $f_B(x)$ and their corresponding $95\%$ confidence intervals are provided in Appendix~\ref{chnader-chhaya}.  

\subsection{Examining the correlation between resource-generating powers for altered measures}

\begin{figure}
\includegraphics[width=8cm]{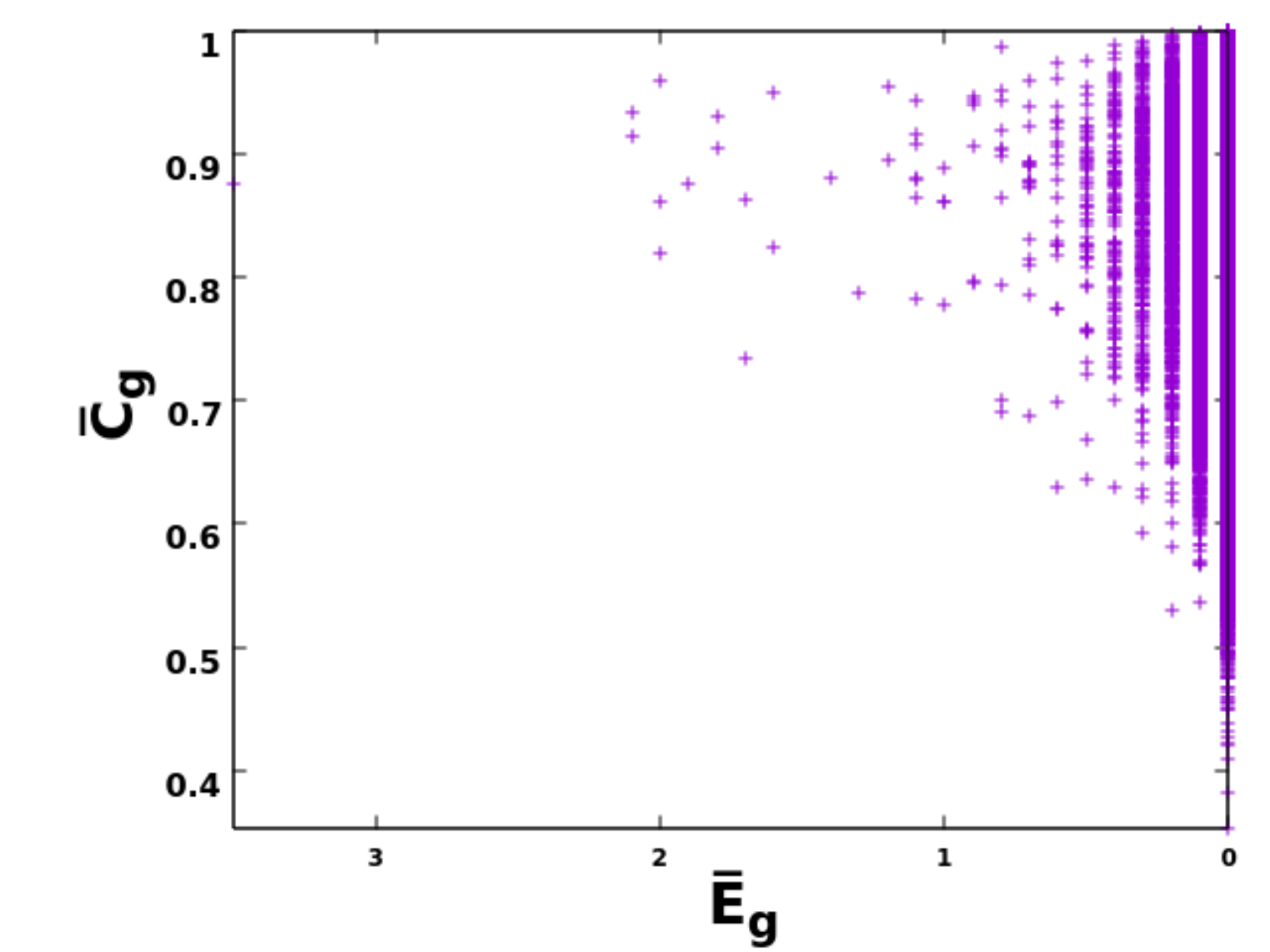}
\caption{Maximum coherence vs maximum entanglement generating power, defined by Eqs.~(\ref{sadher-lau1}) and~(\ref{500miles_l1}), generated by Haar uniform random unitaries.
Here we have presented $\overline{C}_g$ with respect to $\overline{E}_g$ for $13.5 \times 10^{5}$ Haar uniformly chosen two-qubit unitaries.
The quantity plotted along the horizontal axis is in 
ebits and the one along the vertical axis is in bits.
Along the horizontal axis $k$ represents $1-k\times 10^{-5}$, for $k=0,...,3$.}
\label{new_measure}
\end{figure}

We now move over to the discussion of the resource generating capacities of Haar random quantum gates using an altered pair of measures of the resources. Here, we have used a Nielsen-Vidal entanglement monotone and the $l_1$-norm of quantum coherence as measures of entanglement and quantum coherence respectively. The respective resource generating powers, $\overline{E}_g$ and $\overline{C}_g$, given respectively in Eqs.~(\ref{sadher-lau1}) and~(\ref{500miles_l1}), are presented in a scattered plot in Fig.~\ref{new_measure} for $13.5 \times 10^{5}$  
Haar random unitaries. 
Like in the previous investigations, here also, most of the unitaries generate maximum entanglement and maximum coherence. Along the coherence axis, the spread dies out after $0.4$ bits while along the entanglement axis, the spread is very small and most of the points are concentrated between $1-10^{-5}$ and $1$. Hence, in this scenario also, 
the maximum entanglement generating unitaries are the maximum coherence generating ones, and vice versa.

\section{Conclusion}
\label{Sec:6}
There are two main themes of this manuscript. The first is to find the maximum quantum coherence that can be generated by two-qubit unitary gates from pure incoherent states. And the second is to compare this generation with entanglement generation for the same gate from pure unentangled states, for generic two-qubit unitary gates.

With respect to the first theme,  we dealt with the maximum   quantum coherence generated by a general two-qubit unitary operator acting on an incoherent pure state of an arbitrary product  basis of two qubits, as also a generic basis of the same.
We discussed about the best choice of basis for 
%
obtaining the maximum coherence for a fixed unitary, and about the nature of the quantum coherences generated by the unitary when acting on different elements of the basis. The work of Kraus and Cirac~\cite{Kraus} had considered the parallel problem for entanglement generation, and had identified a class of two-qubit unitaries, that they called ``non-local'' unitaries. The same was referred to as the ``Cartan kernel part'' in Khaneja and Glaser~\cite{Khaneja}. It is this part of the whole unitary that is responsible for entanglement generation, provided there is a certain arbitrariness present in the input. We performed the analysis for the 
general two-qubit unitaries, as even local unitaries can generate quantum coherence.
We also analyzed the quantum coherence generating powers of unitaries that can create maximal entanglement, and in parallel, the entanglement power of maximally quantum coherence generating unitaries.


Finally, for the second theme, we considered the correlation between entanglement and quantum coherence generations for a generic two-qubit unitary gate, which we generated Haar uniformly by using the Ginibre ensemble. We found that the high entanglement generating unitaries are also typically high quantum coherence generating ones, and conversely, the unitaries that generate high quantum coherence also typically generate high entanglement. It should be noted however that there is an inherent asymmetry between the entanglement and quantum coherence generations. In particular, we analyzed the profile of the relative frequency of  unitaries that can generate maximal entanglement to have the ability to create a given amount of quantum coherence, and found that it can be well-described by the beta distribution. Role-reversal between entanglement and quantum coherence leads one to obtain a steeper curve, but still describable by the beta distribution, albeit with a different set of distribution parameters.

Relations between different resource theories may help us to understand a general structure among resource theories in quantum systems. 
Entanglement and quantum coherence are the ubiquitous resources in quantum technologies. The connection uncovered here between entanglement and quantum coherence generation for generic two-qubit global unitaries can potentially lead to 
fundamental inter-relations between these resources, as also to fresh applications and towards optimized usage of these resources in quantum devices. The interrelations between different resources of the same quantum device could potentially be of fundamental importance in utilizing these resources in an optimal way. A device that is known to be significantly good for generating a certain resource may not be so for generating another. We however show that the two quintessential quantum resources, viz. entanglement and quantum coherence, are related, and once we know that a device can generate a significant amount of entanglement, then we can be sure - with high probability - that the same can generate a significant amount of quantum coherence, and vice versa. 


\acknowledgements 

We acknowledge computations performed using Armadillo~\cite{Sanderson,Sanderson1},
NLOPT~\cite{NLOPT}
 (Direct-L~\cite{Gablonsky}, ISRES~\cite{Runarsson},
  Cobyla~\cite{Powell}),
   and QIClib~\cite{QIClib}
   on the cluster computing facility of the Harish-Chandra Research Institute, India. We also acknowledge partial support from the Department of Science and Technology, Government of India through the QuEST grant (grant number DST/ICPS/QUST/Theme-3/2019/120).

\appendix
\section{Beta distribution and the fitting parameters for panels \((c)\) and \((d)\) of Fig.~\ref{fig7} and for panel $(b)$ of Fig.~\ref{ent_bas_freq}} 
\label{chnader-chhaya}

The beta distribution is a probability density function  given by 
\begin{equation}
    B(x; \alpha_B, \beta_B)= \frac{\Gamma(\alpha_B+\beta_B)}{\Gamma(\alpha_B)\Gamma(\beta_B)} (x-x_0)^{\alpha_B-1} [1-(x-x_0)]^{\beta_B-1},
\end{equation}
for \(x\in[0,1]\), and \(\alpha_B>0\), \(\beta_B>0\). \(x_0\) is a real number. 
For panel \((c)\) of Fig.~\ref{fig7}, 
the best fit values of the parameters of $f_B(x)$ and their respective
95\% confidence intervals, within the nonlinear least-square fit method, 
are
\begin{eqnarray}
 \alpha_B&=&21.0539 \quad  (\pm 0.2618), \quad \nonumber \\
 \beta_B&=&0.4694 \quad (\pm 0.0143), \quad \quad \nonumber \\
 d&=&0.0135 \quad (\pm 0.0003 ), \nonumber \\
 x_0&=&0.0022  \quad (\pm 0.0001), \nonumber \\
 h&=&-8.0159 \times 10^{-6} \quad (\pm 1.793 \times 10^{-5}).
 \end{eqnarray}
 The values of the same parameters in Fig.~\ref{fig7}$(d)$ are
\begin{eqnarray}
 \alpha_B&=&9.1513 \quad  (\pm 0.1627), \quad \quad \quad \nonumber \\
 \beta_B&=&0.3766 \quad (\pm 0.0097),\nonumber \\
 d&=&0.0019 \quad (\pm 3.049 \times 10^{-5}),\nonumber \\
 x_0&=&6.8175 \times 10^{-5} \quad (\pm 5.969 \times 10^{-6}), \nonumber \\
 h&=&9.6372 \times 10^{-6} \quad (\pm 5.78 \times 10^{-6}).
 \end{eqnarray}
The following are the values of the same parameters in Fig.~\ref{ent_bas_freq}(b):
  \begin{eqnarray}
  \alpha_B&=&8.9527 \quad  (\pm 0.1871), \quad \quad \quad \nonumber \\
 \beta_B&=&0.3949 \quad (\pm 0.0106),\nonumber \\
 d&=&0.0028 \quad (\pm 4.656 \times 10^{-5}),\nonumber \\
 x_0&=&7.2652 \times 10^{-5} \quad (\pm 7.066 \times 10^{-6}), \nonumber \\
 h&=&5.0638 \times 10^{-5} \quad (\pm 1.506 \times 10^{-5}).
  \end{eqnarray}


\begin{thebibliography}{100}
\bibitem{Plenio1} M. B. Plenio and S. Virmani, \textit{An Introduction to entanglement measures},  Quant. Inf. Comput. \textbf{7}, 1 (2007). 
\bibitem{Horodecki} R. Horodecki, P. Horodecki, M. Horodecki and K. Horodecki, \textit{Quantum entanglement}, Rev. Mod. Phys. \textbf{81}, 865 (2009).
\bibitem{Toth} O. G{\"u}hne and G. T{\'o}th, \textit{Entanglement detection}, Physics Reports \textbf{474}, 1 (2009).
\bibitem{Lewenstein} S. Das, T. Chanda, M. Lewenstein, A. Sanpera, A. Sen(De) and U. Sen, \textit{The separability versus entanglement problem}, in \emph{Quantum Information}, edited by D. Bru{\ss} and G. Leuchs (Wiley, Weinheim, 2019), chapter 8.

\bibitem{Bennett1} C. H. Bennett, G. Brassard, C. Cr\'{e}peau, R. Jozsa, A. Peres and W. K. Wootters, \textit{Teleporting an unknown quantum state via dual classical and Einstein-Podolsky-Rosen channels}, Phys. Rev. Lett. \textbf{70}, 1895 (1993).
\bibitem{Pirandola} S. Pirandola, J. Eisert, C. Weedbrook, A. Furusawa and S. L. Braunstein, \textit{Advances in quantum teleportation},  Nat. Photonics \textbf{9}, 641 (2015).
\bibitem{Liu} T. Liu, 
\emph{The Applications and Challenges of Quantum Teleportation},
J. Phys.: Conf. Ser. \textbf{1634}, 012089 (2020).
\bibitem{Gisin} N. Gisin, G. Ribordy, W. Tittel, H. Zbinden, \textit{Quantum Cryptography}, Rev. Mod. Phys. \textbf{74}, 145 (2002).
\bibitem{Pirandola1} S. Pirandola, U. L. Andersen, L. Banchi, M. Berta, D. Bunandar, R. Colbeck, D. Englund, T. Gehring, C. Lupo, C. Ottaviani, J. Pereira, M. Razavi, J. S. Shaari, M. Tomamichel, V. C. Usenko, G. Vallone, P. Villoresi and P. Wallden, \textit{Advances in Quantum Cryptography}, Adv. Opt. Photon. \textbf{12}, 1012 (2020).
\bibitem{Portmann} C. Portmann and R. Renner, \textit{Security in Quantum Cryptography}, 	arXiv:2102.00021.
\bibitem{Wiesner}C. H. Bennett, S. J. Wiesner, \textit{Communication via one- and two-particle operators on Einstein-Podolsky-Rosen states}, Phys. Rev. Lett. \textbf{69}, 2881 (1992). 
\bibitem{Li} Y. Guo, B. H. Liu, C. F. Li and G. C.  Guo, \textit{Advances in quantum dense coding}, Advanced Quantum Technologies, \textbf{2}, 1900011, (2019).
\bibitem{Aberg} J. {\AA}berg, \textit{Quantifying Superposition}, 	arXiv:quant-ph/0612146.
\bibitem{Plenio} T. Baumgratz, M. Cramer and M. B. Plenio, \textit{Quantifying Coherence}, Phys Rev Lett, \textbf{113}, 140401 (2014).

\bibitem{Winter} A. Winter and D. Yang, \textit{Operational Resource Theory of Coherence}, Phys. Rev. Lett. \textbf{116}, 120404 (2016).


\bibitem{Streltsov} A. Streltsov, G. Adesso and M. B. Plenio, \textit{Quantum Coherence as a Resource}, Rev. Mod. Phys. \textbf{89}, 041003 (2017).

\bibitem{Sreetama} 
T. Theurer, N. Killoran, D. Egloff and M. B. Plenio, \emph{Resource Theory of Superposition}, Phys. Rev. Lett. \textbf{119}, 230401 (2017);
S. Das, C. Mukhopadhyay, S. S. Roy, S. Bhattacharya, A. Sen(De) and U. Sen \textit{Wave-particle duality employing quantum coherence in superposition with non-orthogonal pointers}, J. Phys. A: Math. Theor. \textbf{53}, 115301 (2020).
\bibitem{Chirag} C. Srivastava, S. Das and U. Sen, \textit{Resource theory of quantum coherence with probabilistically non-distinguishable pointers and corresponding wave-particle duality}, Phys. Rev. A \textbf{103}, 022417 (2021).
\bibitem{Ignita} I. Banerjee, K. Sen, C. Srivastava and U. Sen, \textit{Quantum coherence with incomplete set of pointers and corresponding wave-particle duality}, arXiv:2108.05849.

\bibitem{Pires} D. P. Pires, I. A. Silva, E. R. deAzevedo, D. O. Soares-Pinto and J. G. Filgueiras, \textit{Coherence orders, decoherence and quantum metrology}, Phys. Rev. A \textbf{98}, 032101 (2018).
\bibitem{Castellini} A. Castellini, R. Lo Franco, L. Lami, A. Winter, G. Adesso and G. Compagno, \textit{Indistinguishability-enabled coherence for quantum metrology}, Phys. Rev. A \textbf{100}, 012308 (2019).
\bibitem{Zhang} C. Zhang, T. R. Bromley, Y.-F. Huang, H. Cao, W.-M. Lv, B.-H. Liu, C.- F. Li, G.-C. Guo, M. Cianciaruso and G. Adesso, \textit{Demonstrating quantum coherence and metrology that is resilient to transversal noise}, Phys. Rev. Lett. \textbf{123}, 180504 (2019).
\bibitem{Hillery} M. Hillery, \textit{Coherence as a resource in decision problems: The Deutsch-Jozsa algorithm and a variation}, Phys. Rev. A \textbf{93}, 012111 (2016). 
\bibitem{Anand} N. Anand and A. K. Pati, \textit{Coherence and Entanglement Monogamy in the Discrete Analogue of Analog Grover Search}, arXiv:1611.04542.
\bibitem{Shi} H.-L. Shi, S.-Y. Liu, X.- H. Wang, W.-L. Yang, Z.-Y. Yang and H. Fan, \textit{Coherence depletion in the Grover quantum search algorithm}, Phys. Rev. A \textbf{95}, 032307 (2017).
\bibitem{Liu1} Y.-C. Liu, J. Shang and X. Zhang, \textit{Coherence Depletion in Quantum Algorithms}, Entropy \textbf{21}, 260 (2019).
\bibitem{Sen} Shubhalakshmi S and U. Sen, \textit{Noncommutative coherence and quantum phase estimation algorithm}, arXiv:2004.01419.
\bibitem{Xiong} C. Xiong and J. Wu, \textit{Geometric coherence and quantum state discrimination}, J. Phys. A: Math. Theor. \textbf{51}, 414005 (2018).
\bibitem{Kim} S. Kim, L. Li, A. Kumar, C. Xiong, S. Das, U. Sen, A. K. Pati and J. Wu, \textit{Unambiguous quantum state discrimination with quantum coherence}, arXiv:1807.04542.

\bibitem{Zanardi2} P. Zanardi, \textit{Entanglement of Quantum Evolutions}, 	Phys. Rev. A \textbf{63}, 040304 (2001). 
\bibitem{Kraus} B. Kraus and J. I. Cirac, \textit{Optimal Creation of Entanglement Using a Two–Qubit Gate}, Phys. Rev. A \textbf{63}, 062309 (2001).
\bibitem{Moor} F. Verstraete, K. Audenaert and B. D. Moor, \textit{Maximally entangled mixed states of two qubits}, Phys. Rev. A \textbf{64}, 012316 (2001).
\bibitem{Zyczkowski} K. {\.Z}yczkowski, P. Horodecki, M. Horodecki and R. Horodecki, \textit{Dynamics of quantum entanglement}, Phys. Rev. A \textbf{65}, 012101 (2002).
\bibitem{Leifer} M. S. Leifer, L. Henderson and N. Linden, \textit{Optimal entanglement generation from quantum operations}, Phys. Rev. A \textbf{67}, 012306 (2003).
\bibitem{Sanders} X. Wang, B. C. Sanders and D. W. Berry, \textit{Entangling power and operator entanglement in qudit systems}, Phys. Rev. A \textbf{67}, 042323 (2003). 
\bibitem{Dodd} P. J.Dodd and J. J.Halliwell, \textit{Disentanglement and decoherence by open system dynamics}, Phys. Rev. A \textbf{69}, 052105 (2004).
\bibitem{Carvalho} A. R. R. Carvalho, F. Mintert and A. Buchleitner, \textit{Decoherence and multipartite entanglement}, Phys. Rev. Lett. \textbf{93}, 230501 (2004).
\bibitem{Dur} W. D\"{u}r and H.-J. Briegel, \textit{Stability of macroscopic entanglement under decoherence}, Phys. Rev. Lett. \textbf{92}, 180403 (2004).
\bibitem{Almeida} M. P. Almeida, F. De Melo, M. Hor-Meyll, A. Salles, S. P. Walborn, P. H. Souto Ribeiro and L. Davidovich, \textit{Environment-induced sudden death of entanglement}, Science \textbf{316}, 579 (2007).
\bibitem{Carvalho1} A. R. R. Carvalho, M. Busse, O. Brodier, C. Viviescas and A. Buchleitner, \textit{Optimal dynamical characterization of entanglement}, Phys. Rev. Lett. \textbf{98}, 190501 (2007).
\bibitem{Bao} H. Bao-Lin and D. Yao-Min, \textit{Entanglement Capacity of Two-Qubit Unitary Operator with the Help of Auxiliary System},  Commun. Theor. Phys. \textbf{47}, 1029 (2007).
\bibitem{Konrad} T. Konrad, F. de Melo, M. Tiersch, C. Kasztelan, A. Arag\~{a}o and A. Buchleitner, \textit{Evolution equation for quantum entanglement}, Nature Physics \textbf{4}, 99 (2008).
\bibitem{Cohen} S. M. Cohen, \textit{All maximally entangling unitary operators}, Phys. Rev. A \textbf{84}, 052308 (2011).
\bibitem{Bennett_arxiv} C. H. Bennett, A. W. Harrow, D. W. Leung, J. A. Smolin, \textit{On the capacities of bipartite Hamiltonians and unitary gates}, IEEE Trans. Inf. Theory, \textbf{49}, (2003).

\bibitem{Frowis} F. Fr\"{o}wis, \textit{Kind of entanglement that speeds up quantum evolution}, Phys. Rev. A \textbf{85}, 052127 (2012).


\bibitem{Chefles} A. Chefles, \textit{Entangling capacity and distinguishability of two-qubit unitary operators}, Phys. Rev. A \textbf{72}, 042332 (2005).
\bibitem{Ishizaka} S. Ishizaka and T. Hiroshima, \textit{Maximally entangled mixed states under non-local unitary operations in two qubits}, Phys. Rev. A \textbf{62}, 022310 (2000).
\bibitem{Plastino} J. Batle, A. R. Plastino, M. Casas and A. Plastino, \textit{On the distribution of entanglement changes produced by unitary operations}, Phys. Lett. A \textbf{307}, 253 (2003). 
\bibitem{Ye} M. Ye, D. Sun, Y. Zhang and G. Guo, \textit{Entanglement-changing power of two-qubit unitary operations}, Phys. Rev. A \textbf{70}, 022326 (2004).
\bibitem{Zanardi1} X. Wang and P. Zanardi, \textit{Quantum entanglement of unitary operators on bipartite systems} Phys. Rev. A \textbf{66}, 044303 (2002).


\bibitem{Oppenheim_Horodecki}J. Oppenheim, K. Horodecki, M. Horodecki, P. Horodecki
and R. Horodecki, \textit{Mutually exclusive aspects of information carried by physical systems: Complementarity between local and non-local information}, Phys. Rev. A \textbf{68}, 022307 (2003).



\bibitem{Ritsch} J. K. Asb\'{o}th, J. Calsamiglia and H. Ritsch, \textit{Computable Measure of Nonclassicality for Light}, Phys. Rev. Lett. \textbf{94}, 173602 (2005).

\bibitem {Yao} Y. Yao, X. Xiao, L. Ge and  C. P. Sun, \textit{Quantum coherence in multipartite systems}, Phys. Rev. A \textbf{92}, 022112 (2015).

\bibitem{Bera} A. Streltsov, U. Singh, H. S. Dhar, M. N. Bera and G. Adesso, \textit{Measuring Quantum Coherence with Entanglement}, Phys. Rev. Lett.
\textbf{115}, 020403 (2015).

\bibitem{Fan} Z. Xi, Y. Li and H. Fan \textit {Quantum coherence and correlations in quantum system}, Sci. Rep. \textbf{5}, 10922 (2015).

\bibitem{Chitambar2} E. Chitambar and M.-H. Hsieh, \textit{Relating the Resource Theories of Entanglement and Quantum Coherence}, Phys. Rev. Lett. \textbf{117}, 020402 (2016).



\bibitem{Plenio4} N. Killoran, F. E. S. Steinhoff and M. B. Plenio, \textit{Converting Nonclassicality into Entanglement}, Phys. Rev. Lett. \textbf{116}, 080402 (2016).



\bibitem{Bera1} A. Streltsov, E. Chitambar, S. Rana, M. N. Bera, A. Winter and M. Lewenstein, \textit{Entanglement and Coherence in Quantum State Merging}, Phys. Rev. Lett. \textbf{116}, 240405 (2016).

\bibitem{Qi} X. Qi, T. Gao and F. Yan, \textit{Measuring Coherence with Entanglement Concurrence}, J. Phys. A: Math. Theor. \textbf{50}, 285301 (2017).

\bibitem{Vedral} H. Zhu, Z. Ma, Z. Cao, S. Fei and V. Vedral, \textit{Operational one-to-one mapping between coherence and entanglement measures}, Phys. Rev. A \textbf{96}, 032316 (2017). 

\bibitem{Chin} S. Chin, \textit{Coherence number as a discrete quantum resource}, Phys. Rev. A \textbf{96}, 042336 (2017).


\bibitem{Zhu} H. Zhu, M. Hayashi and L. Chen
\textit{Axiomatic and operational connections between the l1-norm of coherence and negativity}, 	Phys. Rev. A. \textbf{97}, 022342 (2018).

\bibitem{Plenio_Egloff} D. Egloff, J. M. Matera, T. Theurer, and M. B. Plenio, \textit{Of Local Operations and Physical Wires}, Phys. Rev. X \textbf{8}, 031005 (2018).















\bibitem{Kraemer} L. Kraemer and L. del Rio, \textit{Currencies in Resource Theories}, Entropy \textbf{23}, 755 (2021).




\bibitem{batayan} A. Mekala and U. Sen, \emph{All entangled states are quantum coherent with locally distinguishable pointers}, 
Phys. Rev. A \textbf{104}, L050402 (2021).



\bibitem{Wu} K. Bu, A. Kumar and J. Wu, \textit{Bell-type inequality in quantum coherence theory as an entanglement witness}, arXiv:1603.06322.
\bibitem{Pan} L. Qiu, Z. Liu and F. Pan, \textit{Tripartite Bell-type inequalities for measures of quantum coherence and skew information}, arXiv:1610.07237.
\bibitem{Pati1} D. Mondal, T. Pramanik and A. K. Pati, \textit{non-local advantage of quantum coherence}, Phys. Rev. A \textbf{95}, 010301 (2017).

\bibitem{Titas} T. Chanda and S. Bhattacharya \textit{Delineating incoherent non-Markovian dynamics using quantum coherence},  Annals of Physics \textbf{366}, 1 (2016).
\bibitem{Pati2}S. Bhattacharya, S. Banerjee and A. K. Pati, \textit{Evolution of coherence and non-classicality under global environmental interaction}, Quantum Inf. Process. \textbf{17}, 236 (2018).
\bibitem{Huang} Z. Huang and H. Situ, \textit{Optimal Protection of Quantum Coherence in Noisy Environment}, International Journal of Theoretical Physics \textbf{56}, 503 (2017).
\bibitem{Cakmak} B. Cakmak, M. Pezzutto, M. Paternostro and \"{O}. E. M\"{u}stecaplio\u{g}lu, \textit{Non-Markovianity, coherence and system-environment correlations in a long-range collision model}, Phys. Rev. A \textbf{96}, 022109 (2017).
\bibitem{Pati3} C. Mukhopadhyay, S. Bhattacharya, A. Misra and A. K. Pati, \textit{Dynamics and thermodynamics of a central spin immersed in a spin bath}, Phys. Rev. A \textbf{96}, 052125 (2017).
\bibitem{Borji} A. Mortezapour, M. Ahmadi Borji and R. Lo Franco, \textit{Non-Markovianity and coherence of a moving qubit inside a leaky cavity}, Open Systems and Information Dynamics \textbf{24},   1740006 (2017). 
\bibitem{Yadin} J. Ma, B. Yadin, D. Girolami, V. Vedral and M. Gu, \textit{Converting Coherence to Quantum Correlations}, Phys. Rev. Lett. \textbf{116}, 160407 (2016).
\bibitem{Matera} J. M. Matera, D. Egloff, N. Killoran and M. B. Plenio, \textit{Coherent control of quantum sys- tems as a resource theory}, Quantum Science and Technology \textbf{1}, 01LT01 (2016).
\bibitem{Goswami} Y. Guo and S. Goswami, \textit{Discordlike correlation of bipartite coherence}, Phys. Rev. A. \textbf{95}, 062340 (2017).
\bibitem{Berry} D. W. Berry and B. C. Sanders, \textit{Relation between classical communication capacity and entanglement capability for two-qubit
unitary operations}, Phys. Rev. A \textbf{68}, 032312 (2003).
\bibitem{Pati} A. Misra, U. Singh, S. Bhattacharya and A. K. Pati, \textit{Energy cost of creating quantum coherence}, Phys. Rev. A \textbf{93}, 052335 (2016).


\bibitem{Lin} L. Zhang, Z. Ma, Z. Chen and S. M. Fei, \textit{Coherence generating power of unitary transformations via probabilistic average}, Quantum Information Processing \textbf{17}, 186 (2018).

\bibitem{Zanardi} G. Styliaris, L. C. Venuti and P. Zanardi, \textit{Coherence-generating power of quantum dephasing processes}, Phys. Rev. A \textbf{97}, 032304  (2018).
\bibitem{Smolin} C. H. Bennett, D. DiVincenzo, J. A. Smolin and W. K. Wootters, \textit{Mixed State Entanglement and Quantum Error Correction}, Phys.  Rev.  A  \textbf{54}, 3824 (1996).
\bibitem{Rains} E. M. Rains, \textit{Rigorous treatment of distillable entanglement}, Phys. Rev.  A  \textbf{60}, 173 (1999).
\bibitem{Hayden} P. M. Hayden, M. Horodecki and B. M. Terhal, \textit{The asymptotic entanglement cost of preparing a quantum state}, J. Phys. A: Math. Gen. \textbf{34}, 6891 (2001).
\bibitem{Bennett5} C. H. Bennett, H. J. Bernstein, S. Popescu and B. Schumacher, \textit{Concentrating Partial Entanglement by Local Operations}, Phys. Rev. A \textbf{53}, 2046 (1996).

\bibitem{Nielsen} M. A. Nielsen, \textit{Conditions for a Class of Entanglement Transformations}, Phys. Rev. Lett. \textbf{83}, 436 (1999).
\bibitem{Vidal} Guifré Vidal, \textit{Entanglement of Pure States for a Single Copy},
Phys. Rev. Lett. \textbf{83}, 1046 (1999).
\bibitem{Hardy} L. Hardy, \textit{Method of areas for manipulating the entanglement properties of one copy of a two-particle pure entangled state},
Phys. Rev. A \textbf{60}, 1912 (1999).
\bibitem{Jonathan_Plenio} D. Jonathan and M. B. Plenio, \textit{Minimal Conditions for Local Pure-State Entanglement Manipulation}, 
Phys. Rev. Lett. \textbf{83}, 1455 (1999); Erratum Phys. Rev. Lett. \textbf{84}, 4781 (2000).
\bibitem{Vidal1} Guifré Vidal, \textit{Entanglement monotones}, J. Mod. Opt. \textbf{47}, 2000 (2009).

\bibitem{Winter2} K. B. Dana, M. G. Díaz, M. Mejatty, and A. Winter, \textit{Resource theory of coherence: Beyond states}, Phys. Rev. A \textbf{95}, 062327 (2017).




\bibitem{Walgate} J. Walgate and L. Hardy, \textit{non-locality, Asymmetry, and Distinguishing Bipartite States}, Phys. Rev. Lett. \textbf{89}, 147901 (2002).

\bibitem{Khaneja} N. Khaneja and S. Glaser, \emph{Cartan Decomposition of} \(\text{SU}(2^n)\), \emph{Constructive Controllability of Spin systems and Universal Quantum Computing}, arXiv:quant-ph/0010100.

\bibitem{Hill} S. Hill and W. K. Wootters, \textit{Entanglement of a Pair of Quantum Bits}, Phys. Rev. Lett. \textbf{78}, 5022 (1997).
\bibitem{Wootters} W. K. Wootters, \textit{Entanglement of Formation of an Arbitrary State of Two Qubits}, Phys. Rev. Lett. \textbf{80}, 2245 (1998).

Phys. Rev. Lett. \textbf{83}, 1046 (1999).

\bibitem{NLOPT} S. G. Johnson, \textit{The NLopt nonlinear-optimization package}, \url{http://github.com/stevengj/nlopt}.

\bibitem{Ginibre} M. Ozols, \textit{How to generate a random unitary matrix} (2009). Available at \url{http://home.lu.lv/~sd20008/papers/essays/Random%20unitary%20%5Bpaper%5D.pdf}.

\bibitem{banalata} D. M. Bates and D. G. Watts, \emph{Nonlinear Regression Analysis
and Its Applications} (Wiley, New York, 1988);
W. H. Press, S. A. Teukolsky, W. T. Vetterling and B. P.
Flannery, \emph{Numerical Recipes: The Art of Scientific Computing}
(Cambridge University Press, Cambridge, 2007); A. Sehrawat, C. Srivastava and U. Sen, \emph{Dynamical phase transitions in the fully connected quantum Ising model: Time period and critical time},
Phys. Rev. B \textbf{104}, 085105  (2021), Appendix E. 
\bibitem{Sanderson}  C. Sanderson and R. Curtin, \textit{Armadillo: a template-based C++ library for linear
algebra}, Journal of Open Source Software 1, 26 (2016).
\bibitem{Sanderson1} C. Sanderson and R. Curtin, Lecture Notes in Computer Science (LNCS) 10931, 422 (2018).
\bibitem{Gablonsky} J. M. Gablonsky and C. T. Kelley, \textit{A locally-biased form of the DIRECT algorithm}, J. Global Optimization, \textbf{21},27 (2001).
\bibitem{Runarsson} T. P. Runarsson and X. Yao, \textit{Search biases in constrained evolutionary optimization}, IEEE Trans. on Systems, Man, and Cybernetics Part C: Applications and Reviews, \textbf{35}, 233 (2005).
\bibitem{Powell} M. J. D. Powell, \textit{Direct search algorithms for optimization calculations}, Acta Numerica \textbf{7}, 336 (1998).
\bibitem{QIClib} T. Chanda, \emph{QIClib}, \url{https://titaschanda.github.io/QIClib}.


\end{thebibliography}
\end{document}